# From Crop and Energy Data to Optimal Lighting Scheduling: A Surrogate-Based MILP Framework for Vertical Farming

Francesco Ceccanti[1]*, Andrea Baccioli[1], Aldo Bischi[1]

1 Department of Energy, Systems, Territory and Construction Engineering
Università di Pisa, Pisa, Italy

*Corresponding Author: francesco.ceccanti@phd.unipi.it (F. Ceccanti)

**Abstract**
Vertical farming enables high productivity and stable crop production under controlled conditions, but its economic feasibility is still constrained by the large electricity demand of artificial lighting and climate control. This study develops a surrogate-based optimization framework for cost-minimizing lighting management in hydroponic vertical farms. One of the contributions is a general procedure to convert crop–energy response data into relationships suitable for mathematical programming. The proposed methodology can be applied to either monitored experimental datasets or outputs from validated dynamic models, provided that the relevant ranges of control variables, crop development stages, and energy responses are covered.
In this work, the framework is applied to lettuce cultivation in an industrial-scale vertical farm in northern Italy. Synthetic response data generated by a verified agri-energy model are used to derive surrogate relationships for fresh biomass growth, leaf area index evolution, and lighting, heating, and cooling electricity demand as functions of photosynthetic photon flux density, crop developmental stage, and outdoor temperature. Different temporal aggregation levels are analyzed to assess the trade-off between surrogate accuracy, scheduling flexibility, and computational compactness.
The surrogate model reproduced the main crop and energy trends of the physical reference model with sufficient accuracy for optimization. Compared with fixed-lighting benchmarks, the optimized schedules reduced the specific electric energy consumption by up to 15.9% and the specific electricity cost by up to 17.8% in the studied scenario, while reaching the target harvest weight. The results show that moderate light intensities combined with flexible photoperiods are more cost-effective than high-intensity lighting under the investigated conditions. The proposed framework provides a tractable and generalizable approach for integrating crop development, energy demand, and electricity price variability into vertical farming lighting strategies.

**Keywords:**
Controlled environment agriculture; Vertical farming; MILP; Optimization; Lighting Schedule; Energy System; Energy Consumption

Nomenclature

| **Acronyms** | | | **Subscripts** | |
|---|---|---|---|---|
| **AHU** | Air Handling Unit | $[-]$ | **ahu** | Air Handling Unit |
| **CEA** | Controlled Environment Agriculture | $[-]$ | **$CO_2$** | Carbon Dioxide |
| **COP** | Coefficient of Performance | $[kW_{th}\ kW_{el}^{-1}]$ | **dm** | Dry Matter |
| **DLI** | Daily Light Integral | $[\mu mol\ m^{-2}\ day^{-1}]$ | **env** | Envelope |
| **HVAC** | Heating, Ventilation and Air Conditioning | $[-]$ | **eva** | Evaporation |
| **LAI** | Leaf Area Index | $[{m_{leaf}}^2\ {m_{soil}}^{-2}]$ | **ext** | External |
| **LUE** | Light Use Efficiency | $[g\ \mu mol^{-1}]$ | **fm** | Fresh Matter |
| **MILP** | Mixed Integer Linear Programming | $[-]$ | **H/C** | Heating&Cooling Systems |
| **PPE** | Photosynthetic Photon Efficiency | $[\mu mol\ J^{-1}]$ | **H** | Heating |
| **PPFD** | Photosynthetic Photon Flux Density | $[\mu mol\ m^{-2}\ s^{-1}]$ | **hum** | Humidification System |
| **RMSE** | Root Mean Square Error | $[-]$ | **hw** | Hot water |
| **SEEC** | Specific Electric Energy Consumption | $[kWh\ kg^{-1}]$ | **nom** | Nominal |
| **SC** | Specific Cost | $[€\ kg^{-1}]$ | | |
| **SLA** | Specific Leaf Area | $[cm^2\ g^{-1}]$ | | |
| **VF** | Vertical Farm | $[-]$ | | |
| **Symbols** | | | | |
| $\boldsymbol{A}$ | Area | $[m^2]$ | | |
| $\boldsymbol{c_p}$ | Specific Heat Capacity | $[J\ kg^{-1}\ K^{-1}]$ | | |
| $\boldsymbol{d}$ | Density | $[\$]$ | | |
| $\boldsymbol{DM}$ | Dry Matter | $[g\ m^{-2}]$ | | |
| $\boldsymbol{ET}$ | Evapotranspiration | $[g\ m^{-2}\ s^{-1}]$ | | |
| $\boldsymbol{FM}$ | Fresh Matter | $[-]$ | | |
| $\boldsymbol{k}$ | Light Extinction Coefficient | $[-]$ | | |
| $\boldsymbol{Q}$ | Thermal Power | $[kW]$ | | |
| $\boldsymbol{r_a}$ | Aerodynamic resistance | $[s\ m^{-1}]$ | | |
| $\boldsymbol{r_s}$ | Stomatal resistance | $[s\ m^{-1}]$ | | |
| $\boldsymbol{RF}$ | Root Factor | $[-]$ | | |
| $\boldsymbol{T}$ | Temperature | $[°C]$ | | |
| $\boldsymbol{V}$ | Volume | $[m^3]$ | | |
| $\boldsymbol{X}$ | Vapour Concentration | $[g\ m^{-3}]$ | | |
| **Greeks** | | | | |
| $\boldsymbol{\eta}$ | Efficiency | $[-]$ | | |

## 1. Introduction

Fresh horticultural production is particularly exposed to the increasing variability and constraints affecting modern agriculture. Leafy vegetables and other high-value crops require stable environmental conditions to ensure yield, quality, and marketability, yet they are highly sensitive to fluctuations in temperature, solar radiation, water availability, and pest pressure. This vulnerability is becoming more relevant as global agricultural production is projected to increase by 14% over the next decade mainly through productivity gains, while land and water resources are already affected by depletion, degradation, pollution, biodiversity loss, and increasing scarcity [1]. Climate change further intensifies these constraints, as agricultural systems respond strongly to changes in temperature, precipitation, drought, heat stress, and extreme events [2]. At the same time, fresh horticultural crops are characterized by high perishability and short shelf life, making production stability and supply continuity particularly important.

Within this context, controlled environment agriculture (CEA) has gained increasing attention as a way to reduce the dependence of crop production on external climatic conditions and to improve resource use efficiency. Vertical farming is generally considered one of the most advanced forms of controlled environment agriculture, as it relies on crop cultivation in stacked layers within enclosed systems where the plant environment is precisely managed through electric lighting, climate control, and soilless cultivation technologies [3]. In these systems, key growth factors such as light, temperature, relative humidity, carbon dioxide concentration, irrigation, and nutrient supply are tightly regulated, allowing plant growth to be steered more precisely than in open field or conventional greenhouse production [4]. This high level of control enables year-round production, high spatial productivity, and more uniform product quality, making vertical farming particularly relevant for fresh horticultural crops produced close to consumption areas [5]. However, recent literature also emphasizes that vertical farming is currently most suitable for short cycle, high value crops, especially leafy greens and lettuce, whereas staple crops with high biomass demand remain poorly suited to fully indoor systems because of their high energy requirements and limited economic feasibility [6].
Hydroponics is a key enabling technology for vertical farming because it allows crop production without soil through the controlled delivery and recirculation of nutrient solutions. This makes it particularly suitable for stacked indoor systems, where root zone conditions, water use, and nutrient supply must be managed precisely across multiple growing layers. One of its main advantages is the strong reduction in water demand, since hydroponic systems can recycle nutrient solutions and limit losses compared with soil-based cultivation. Recent reviews report that hydroponic farming can reduce water use by up to 90% compared with conventional soil-based systems, making it particularly attractive in contexts affected by water scarcity and resource constraints [7].

The diffusion of vertical farming is closely linked to hydroponic technologies, which represent the largest growing mechanism segment due to their suitability for automation, indoor cultivation, and high-density production [8]. Industrial cases such as Planet Farms in Italy show that hydroponic vertical farms have reached commercial scale [9]. However, recent company closures highlight that profitability remains challenged by high operating costs, making efficiency improvements essential to support wider adoption while maintaining yield and product quality.

Despite the agronomic advantages of hydroponic vertical farming, energy consumption remains one of its main limiting factors. Unlike open field and greenhouse systems, vertical farms fully replace solar radiation with artificial lighting and require continuous climate control to manage temperature, humidity, and heat loads within the growing environment. As a consequence, lighting and HVAC systems account for most of the total energy demand. Arcasi et al. reported that, in lettuce vertical farming, lighting can represent 65 to 85% of total energy use, while HVAC accounts for an additional 10 to 20%, with energy costs varying strongly across climatic and economic contexts [10]. Similarly, recent quantitative assessments indicate that the energy consumption of vertical farms may range from 3.2 to 20 kWh $kg^{-1}$ of fresh product, depending on LED efficiency, HVAC configuration, environmental setpoints, and operating conditions [11]. This high energy requirement directly affects both production cost and environmental performance. Stanghellini and Katzin emphasized that vertical farms can require substantially more electricity than greenhouse or open field systems, challenging the assumption that local indoor production is automatically more sustainable [6]. More recent benchmarking studies further show that productivity must be interpreted together with energy use, water demand, cost, and carbon emissions, since the conditions that maximize yield do not necessarily minimize specific energy consumption or emissions. In particular, photosynthetic photon flux density (PPFD) has been identified as a dominant driver of both lettuce productivity and energy consumption, confirming that lighting management is central to improving the efficiency of vertical farms [12]. These findings indicate that the establishment of hydroponic vertical farming as a viable technique for vegetable production depends on strategies capable of reducing lighting and climate-control costs while preserving yield, quality, and marketability.

Given the dominant contribution of artificial lighting to energy demand in vertical farming, recent research has increasingly shifted from static light recipes toward dynamic lighting management. Conventional indoor production systems typically rely on fixed PPFD levels and predefined photoperiods designed to achieve a target daily lighting integral (DLI). However, this approach does not account for temporal variability in electricity prices, plant developmental stage, or the intrinsic flexibility of plant physiological responses to light. As a result, static lighting strategies are often suboptimal from both an energetic and economic perspective [4]. The agronomic relevance of flexible light intensity is supported by studies showing that lettuce growth and quality are strongly affected by both PPFD and photoperiod. Matysiak et al. tested romaine lettuce under DLI values of 9.2, 11.5, 13.8, and 17.3 mol $m^{-2}$ $d^{-1}$ and reported marked differences in fresh weight, leaf number, head circumference, sensory quality, and image-based classification accuracy, confirming that light dose management strongly affects commercial performance [13]. A recent review on translational photobiology also supports the use of dynamic light regimes in indoor horticulture, highlighting that crop responses to fluctuating light can be translated into lighting strategies designed around plant physiological acclimation and energy efficiency [14].

One of the earliest and most straightforward dynamic approaches is load shifting, where lighting periods are redistributed over time to exploit lower electricity price windows while maintaining the required light and dark intervals for plant growth. Avgoustaki and Xydis demonstrated that shifting electricity demand in indoor vertical farming can reduce lighting costs by 16 to 26% across different months, without negatively affecting crop performance [15]. This approach leverages the fact that many crops can tolerate interruptions in light supply, provided that minimum photoperiod and total daily light requirements are respected.

Beyond shifting the timing of lighting periods, a second line of research focuses on modulating light intensity itself. In these approaches, crops are not exposed to a constant PPFD throughout the photoperiod, but receive the target DLI through variable light intensity patterns. In greenhouse systems, Alaviani et al. formulated an optimal supplemental lighting control problem based only on DLI, showing that LED lighting can be regulated over an extended photoperiod to minimize electricity cost and demand charge while satisfying crop light requirements [16]. In fully indoor vertical farms, where all light is supplied artificially, the same principle can be exploited to redistribute PPFD over time in response to electricity prices and grid requirements. Penuela et al. showed that indoor farming systems dedicated to leafy greens can participate in demand response programs without adversely affecting plant production, achieving savings of 15.34 to 23.03% in complementary energy costs under simulated demand response conditions [17]. This flexibility is enabled by dimmable LED fixtures, which allow real time regulation of light intensity and integration with sensors and control algorithms for adaptive lighting management. More recent work further suggests that the temporal distribution of light can influence light use efficiency (LUE) even when total light input is kept constant. Lin et al. showed that dynamic light intensity increased lettuce shoot fresh weight by 28% and dry weight by 65% compared with constant lighting under the same total light integral, linking the response to improved canopy light interception and growth efficiency [18]. These findings suggest that lighting strategies should not only shift light across the day, but also consider how crop growth changes with PPFD, photoperiod, developmental stage, and physiological acclimation. However, incorporating detailed crop-growth dynamics into energy optimization models is challenging. Time-dependent crop models are typically nonlinear and may also be non-convex, discontinuous, or non-differentiable. In contrast, scheduling problems that balance model detail and computational tractability are commonly formulated using linear or mixed-integer linear programming. These formulations require some degree of model simplification, whose accuracy depends on the adopted piecewise-linear approximation, but they allow the global optimum of the resulting optimization problem to be identified.

Despite the growing body of work on dynamic lighting strategies, a clear gap remains between biologically realistic crop modeling and energy aware optimization in vertical farming. Existing approaches often simplify plant responses through fixed DLI constraints, constant light use efficiency, or empirical biomass relationships, while more detailed mechanistic crop models are generally too complex to be directly embedded in optimization frameworks. Previous studies have optimized lighting schedules using DLI based constraints [16], time variable electricity prices [15], or crop growth models [19], reporting substantial cost reductions, but often with limited representation of crop physiology or product quality responses. Conversely, recent work on dynamic light intensity and plant photobiology has shown that plant responses depend on growth stage, light intensity, and physiological acclimation, but these insights are not always translated into operational energy optimization frameworks [18]. As a result, few studies consider PPFD-dependent light use efficiency, crop developmental stage, and energy consumption within the same decision-making framework as well as the electricity price variability. At the same time, systematic procedures are still needed to identify the minimum set of input–output relationships required to synthesize complex crop–energy responses into tractable surrogate formulations suitable for mathematical programming-based optimization, whether these relationships are derived from experimental datasets or from validated dynamic models.

The main contribution of this study is the development of a computational procedure to derive a compact set of input–output relationships that capture the key nonlinear interactions among crop growth, energy consumption, light intensity, developmental stage, photoperiod, and external temperature. These relationships were formulated as surrogate models suitable for integration into a mathematical programming framework. Specifically, they were piecewise linearized and incorporated into a mixed-integer linear programming formulation (MILP). The procedure is generally applicable to experimental datasets or to synthetic datasets generated by validated crop–energy models, provided that the relevant operating ranges are adequately covered. In this study, it is applied to synthetic data produced by the abovementioned validated dynamic model, previously assessed against experimental agronomic data and through intermodel comparison for energy performance [20].

The resulting surrogate model is embedded in a MILP framework to optimize lighting management over the post transplanting crop cycle. Specifically, PPFD and photoperiod are dynamically selected as functions of electricity prices and crop developmental stage, with the objective of minimizing energy costs while satisfying agronomic constraints. This allows the framework to retain the key biological and energy-related dependencies of the original mechanistic model while remaining computationally tractable.

Overall, the proposed approach enables the quantification of energy and economic savings from adaptive lighting strategies in hydroponic vertical farms, while preserving a biologically grounded representation of crop development. The derived relationships may also support techno-economic assessments requiring simplified yet biologically informed estimates of vertical farm productivity and energy consumption.

Hence, the main scientific contributions of this work are summarized as follows:

- Develop a systematic procedure to derive a surrogate model suitable for mathematical programming optimization from crop–energy response data.
- Derive piecewise linear relationships as first-order design tools for estimating energy consumption and crop productivity in hydroponic vertical farms operating under comparable conditions, supporting techno-economic evaluations.
- Identify optimized lighting schedules, as per state-of-the-art operations research, via Mixed Integer Linear Programming, over the entire crop cycle by adapting both PPFD and photoperiod to time variable electricity price signals and crop state, with the objective of minimizing energy costs under agronomic constraints.

## 2. Methodology

This section describes the procedure used to transform crop–energy response data into a piecewise-linear surrogate formulation suitable for integration into a MILP optimization framework. The overall workflow, summarized in Figure *1*, is designed to identify a compact set of input–output relationships able to represent the main dependencies among crop growth, energy consumption, lighting conditions, crop developmental stage, and external temperature within a tractable optimization framework. In this study, the response data are generated using the AMESim vertical farm model, which provides reference agronomic and energy outputs under different operating conditions. These outputs are then processed through a data conditioning phase, including interpolation, discretization, and grouping, to derive a surrogate model that approximates the original nonlinear behavior. In parallel, the thermal loads are converted into heat-pump electricity consumption using COP-based correlations derived from heat-pump performance data. The resulting electrical consumption, which therefore accounts for the heat pump's operating efficiency, is used as an input to the surrogate model. The consistency of the surrogate model is verified before its integration into the optimization framework. Finally, the linearized model, combined with economic inputs, is embedded in a MILP formulation to identify the optimized lighting schedule. The framework is formulated deterministically, since the limited availability of experimental data for industrial vertical farms does not currently support a reliable probabilistic characterization of crop- and energy-model uncertainties.

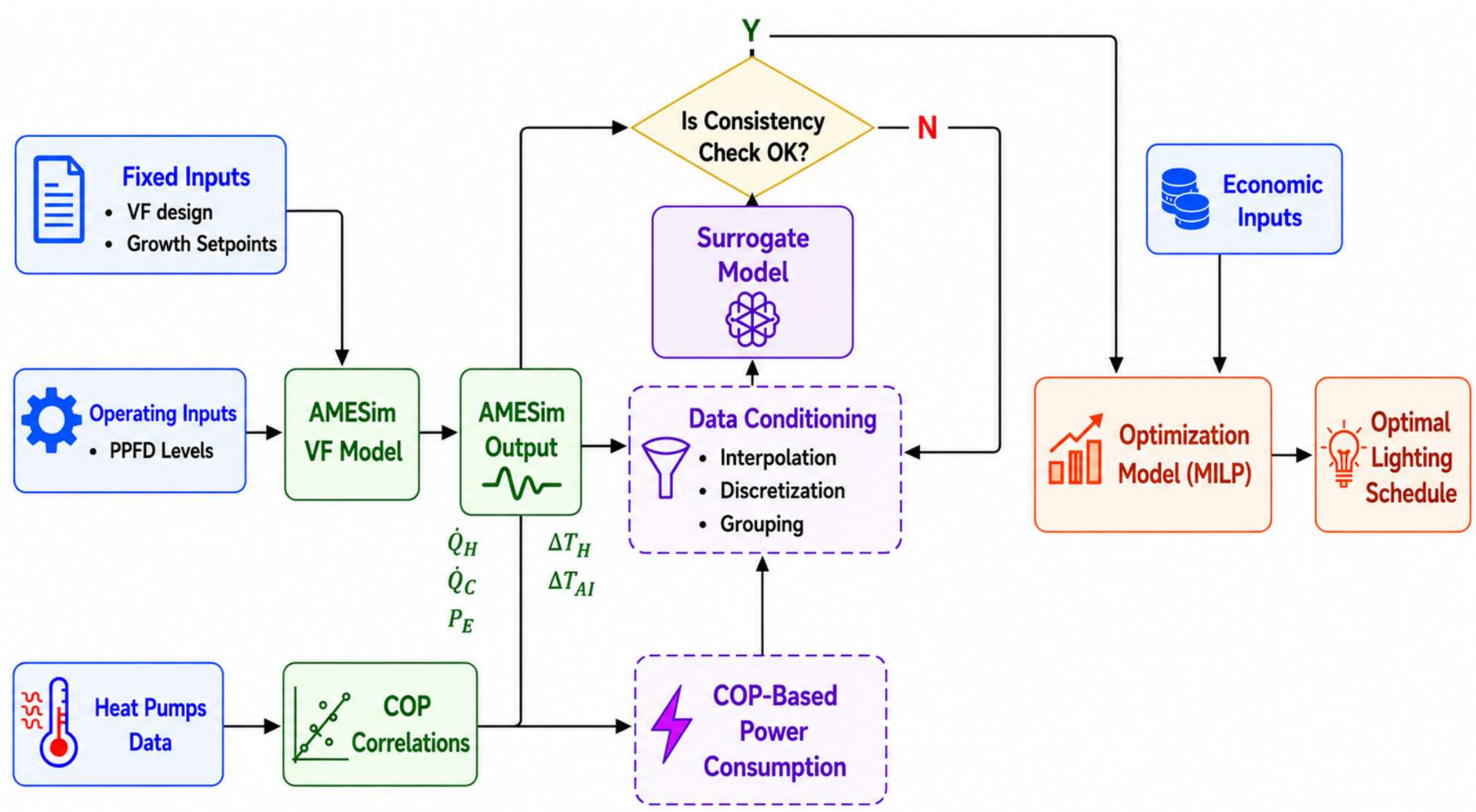

Figure 1: Workflow of the surrogate-based optimization procedure used to convert AMESim crop-energy outputs into a piecewise linear surrogate model and integrate it into a MILP framework for optimized lighting schedule identification.

### 2.1. Reference vertical farm configuration and operating assumptions

The AMESim VF model used in this study represents an industrial-scale vertical farm with a total volume of 6,720 m³, corresponding to a 21 m × 40 m × 8 m facility. This size was selected to describe a system closer to current industrial developments than the smaller scale vertical farms commonly considered in the literature, currently focused on lab scale applications [21]. The cultivation layout consists of racks with 10 tiers, resulting in a total cultivated area of 5,040 m², corresponding to 60% of the overall floor area. Given the high level of envelope insulation, heat losses through the envelope have a relatively low impact on total energy consumption. Under these assumptions, model outputs mainly driven by cultivated area and lighting power can be approximately rescaled through linear relationships for preliminary analyses of facilities with comparable layouts and operating conditions.

Lettuce was selected as the reference crop for the agri-energy model [20]. This crop is among the most widely studied species in vertical farming research and one of the most commonly cultivated in commercial systems, due to its short production cycle, high harvest index, favorable response to artificial lighting, and relatively high market value. Pattison et al. identified leafy vegetables, including lettuce, as particularly suitable crops for vertical farming applications [22]. In the present study, lettuce cultivation was assumed to be hydroponic, consistently with current vertical farming practice, where hydroponics is widely adopted because of its operational simplicity, compatibility with automation, and suitability for soil free indoor production [8]. A planting density of 25 plants $m^{-2}$ was adopted [23].

The growing conditions adopted in this study were selected based on values commonly reported in the literature. A $CO_2$ concentration of 1,200 ppm and a reference photoperiod of 16 $h\,d^{-1}$ were adopted following Carotti et al. [23], who used these conditions for lettuce cultivation in controlled environments. Relative humidity has been shown to have a limited effect on vertical farm performance compared with other environmental variables [24], however, its control remains necessary to avoid excessive moisture conditions and reduce the risk of microbial proliferation on leaves. For this reason, relative humidity setpoints of 75% during the light period and 85% during the dark period were selected [20]. Finally, an indoor air temperature setpoint of 24 °C was adopted, as this value showed favorable results for crop growth in studies evaluating the influence of vertical farm design and operating parameters on system performance [12]. All the parameters adopted in this study are summarized in Appendix 6.1.

Although these setpoints may vary among commercial vertical farms, they fall within typical operating ranges for hydroponic lettuce cultivation. Therefore, the proposed workflow is transferable to other vertical-farming applications, while the resulting surrogate relationships remain specific to the investigated crop and system configuration. Since the adopted operating conditions are representative of hydroponic lettuce production, these relationships may also support preliminary assessments of comparable facilities.

Light intensity was identified as the main driver of both energy consumption and crop growth, with a correlation coefficient of 0.85 [12]. For this reason, PPFD was selected as one of the main optimization variables in this study. Varying PPFD directly affects the electricity required by the LED system and, at the same time, modifies the heat released into the growing environment, thereby influencing the cooling demand. From the agronomic side, PPFD controls the light input available for photosynthesis and therefore affects biomass accumulation, although the relationship between light intensity and growth is nonlinear due to variations in light use efficiency and crop developmental stage.

To characterize these coupled agronomic and energy responses, five PPFD levels were simulated in the agri-energy model [20], namely 100, 200, 300, 400, and 500 $\mu mol\, m^{-2}\, s^{-1}$. This range was selected to cover both low intensity and high intensity lighting conditions commonly considered for lettuce cultivation in controlled environments, allowing the model to capture the nonlinearities associated with different light regimes. The resulting simulations were used to derive response curves linking PPFD to crop growth, lighting electricity demand, and climate control requirements (temperature, relative humidity).

To convert lighting setpoints into electrical and thermal loads, three photosynthetic photon efficacy (PPE) values were considered, equal to 2.0, 2.5, and 3.0 $\mu mol\ J^{-1}$. These values represent different LED efficiency scenarios, from relatively conservative fixtures to more efficient horticultural lighting technologies. Including multiple PPE levels allowed the analysis to account for the impact of lighting technology on both direct electricity consumption and the heat load introduced into the vertical farm environment.

To fully electrify the vertical farm and reduce direct carbon emissions associated with indoor climate control, heating, cooling, and dehumidification loads were supplied by an HVAC system based on two separate electrically driven units. A heat pump was used to provide space heating and post heating within the air handling unit, whereas a chiller was used to provide space cooling and dehumidification. This configuration was adopted because heating and dehumidification loads may occur simultaneously, requiring independent heating and cooling effects.

Both units were modeled as air to water systems, with nominal supply water temperatures of 45 °C for heating and 7 °C for cooling. Based on performance data reported by commercial manufacturers for air to water heat pumps operating around A7/W45 conditions, a nominal COP of 3.4 was assumed for the heating unit. This value is consistent with typical rated performances of air source heat pumps at medium temperature supply levels. For the chiller, a nominal EER of 2.9 was assumed at 7 °C chilled water supply temperature and 35 °C outdoor air temperature, consistently with manufacturer data for air cooled liquid chillers under standard full load rating conditions.

The nominal capacities were selected from preliminary agri-energy simulations performed using the physical model, at the upper bound of the investigated PPFD range, equal to 500 $\mu mol\, m^{-2}\, s^{-1}$. This condition was used as the sizing scenario because it represents the highest lighting power input and therefore the most demanding case in terms of sensible and latent loads. The resulting nominal capacities were 450 kW for the heat pump and 1,100 kW for the chiller.

### 2.2. Reference Physical Models

This section summarizes the physical reference models used to generate and characterize the crop–energy response data from which the piecewise linear surrogate formulation was derived.

#### 2.2.1. AMESim VF Model

The complete set of equations implemented in the AMESim agri energy model is reported and discussed in detail in the reference model paper [20]. In brief, the model combines a mechanistic growth model for the selected crop with an energy model describing the thermal behavior of the vertical farm. The latter accounts for the main sensible and latent heat exchanges occurring within the growing chamber. The heat balance adopted in the model is summarized in Eq. (1).

$$d_{air} c_{p_{air}} V_{room} \frac{\partial T}{\partial t}(t) = Q_{env}(t) + Q_{LED}(t) - Q_{plant}(t) - Q_{eva}(t) + Q_{h/c}(t) - Q_{AHU}(t) - Q_{hum}(t) \quad (1)$$

$Q_{LED}$ is the radiant power supplied by the lighting system, while $Q_{plant}$ represents the fraction intercepted by the crop and converted through photosynthetic processes. The sensible heating or cooling load provided to the room is represented by $Q_{h/c}$, whereas $Q_{AHU}$ accounts for the thermal load associated with air handling, including dehumidification and post heating. Latent heat exchanges are represented by $Q_{eva}$, associated with crop evapotranspiration, and $Q_{hum}$, associated with humidification. Finally, $Q_{env}$ describes heat transfer through the chamber envelope.

Crop growth was described using the validated mechanistic formulation presented in [20]. The model estimates biomass accumulation as a function of air temperature, $CO_2$ concentration, and incident PPFD, through their effects on canopy light interception and light use efficiency (LUE). Dry and fresh matter production were computed at each timestep from the fraction of incident light intercepted by the canopy, described through a Beer-Lambert formulation based on the leaf area index (LAI). In the model, DM and FM denote dry and fresh biomass, respectively, k is the light extinction coefficient, and $A_{crop}$ is the cultivated area. The root factor RF, set to 0.07, represents the fraction of dry biomass allocated to root growth and therefore excluded from the biomass used to update LAI (see Eqs. (2) and (3)). Baseline LUE values were taken from the experimental dataset of Carotti et al. [23] and adjusted for different $CO_2$ concentrations using the data reported in [25]. LAI dynamics were derived from dry matter accumulation through the specific leaf area (SLA) according to Eq. (4). Since leaf morphology changes with light conditions, the dependence of SLA on incident PPFD was explicitly included in the model. The resulting relationship is reported in Appendix 6.2, while further details on the full model formulation and parametrization are provided in the reference model paper [20]. Short-term physiological memory effects, including light acclimation and circadian responses, are not explicitly represented in the presented formulation.

$$\frac{\partial DM}{\partial t}(t) = PPFD(t) \cdot (1 - e^{-k \cdot LAI(t)}) \cdot LUE_{dm} \cdot A_{crop} \cdot (1 - RF) \tag{2}$$

$$\frac{\partial FM}{\partial t}(t) = PPFD(t) \cdot (1 - e^{-k \cdot LAI(t)}) \cdot LUE_{fm} \cdot A_{crop} \tag{3}$$

$$\frac{\partial LAI}{\partial t}(t) = \frac{\partial DM}{\partial t}(t) \cdot SLA(t) \tag{4}$$

The crop growth model was coupled with the crop transpiration model proposed by Graamans et al. [26] to estimate the latent heat load associated with dehumidification. The model was developed for lettuce production in closed plant production systems and describes the relationship between sensible and latent heat exchange and the corresponding vapour flux from the crop canopy. It is based on an adapted Penman Monteith formulation using a big leaf approach, where transpiration is driven by the humidity gradient between the leaf surface and the surrounding air and regulated by surface and aerodynamic resistances. In Eq. (5), $X_a$ is the absolute humidity of the indoor air, $X_s$ is the saturated absolute humidity at leaf temperature, while $r_s$ and $r_a$ denote the surface and aerodynamic resistances, respectively.

Since the original formulation was validated assuming a recommended LAI of 3, an effective leaf area index ($LAI_{eff}$) was introduced in this work to couple transpiration with the crop growth model. $LAI_{eff}$ was set equal to the LAI calculated by the crop growth model up to a maximum value of 3, which was used to represent a fully developed lettuce canopy in the model implementation. The resulting transpiration rate was then used to estimate the moisture removal requirement and the corresponding dehumidification load. Further details on the assumptions and equations implemented in AMESim to compute crop transpiration at each timestep are reported in Appendix 6.3.

$$ET(t) = LAI_{eff}(t) \cdot \frac{(X_s(t) - X_a(t))}{r_s(t) + r_a} \quad (5)$$

To assess the reliability of the AMESim implementation, an intermodel comparison was carried out against the dynamic vertical farm model developed by Talbot et al [24]. The comparison was performed over a range of indoor temperature setpoints and PPFD levels, considering both the annual energy load and the specific energy consumption per unit of harvested biomass, as shown in Figure *2*.

Figure *2*(a) compares the annual energy load predicted by the present model with that obtained from the reference model. Most points fall within the ±15% deviation range, indicating a satisfactory agreement across the investigated operating conditions. The deviation tends to increase at higher PPFD levels, where the present model predicts higher annual energy loads. This trend is likely influenced not only by differences in the energy model itself, but also by discrepancies in the crop growth formulation, since higher PPFD values amplify differences in predicted biomass accumulation and cultivation dynamics. This interpretation is supported by Figure *2*(b), where the comparison is expressed in terms of specific energy consumption. After normalizing energy demand by harvested biomass, the agreement between the two models improves, with most points lying within the ±10% range. The lower deviation observed for this indicator suggests that the energy balance and HVAC load estimation are broadly consistent with the reference model, while part of the discrepancy in annual energy load can be attributed to differences in biomass prediction. Overall, the intermodel comparison supports the suitability of the AMESim agri-energy model [20] for evaluating energy demand and specific energy performance under different lighting and temperature conditions.

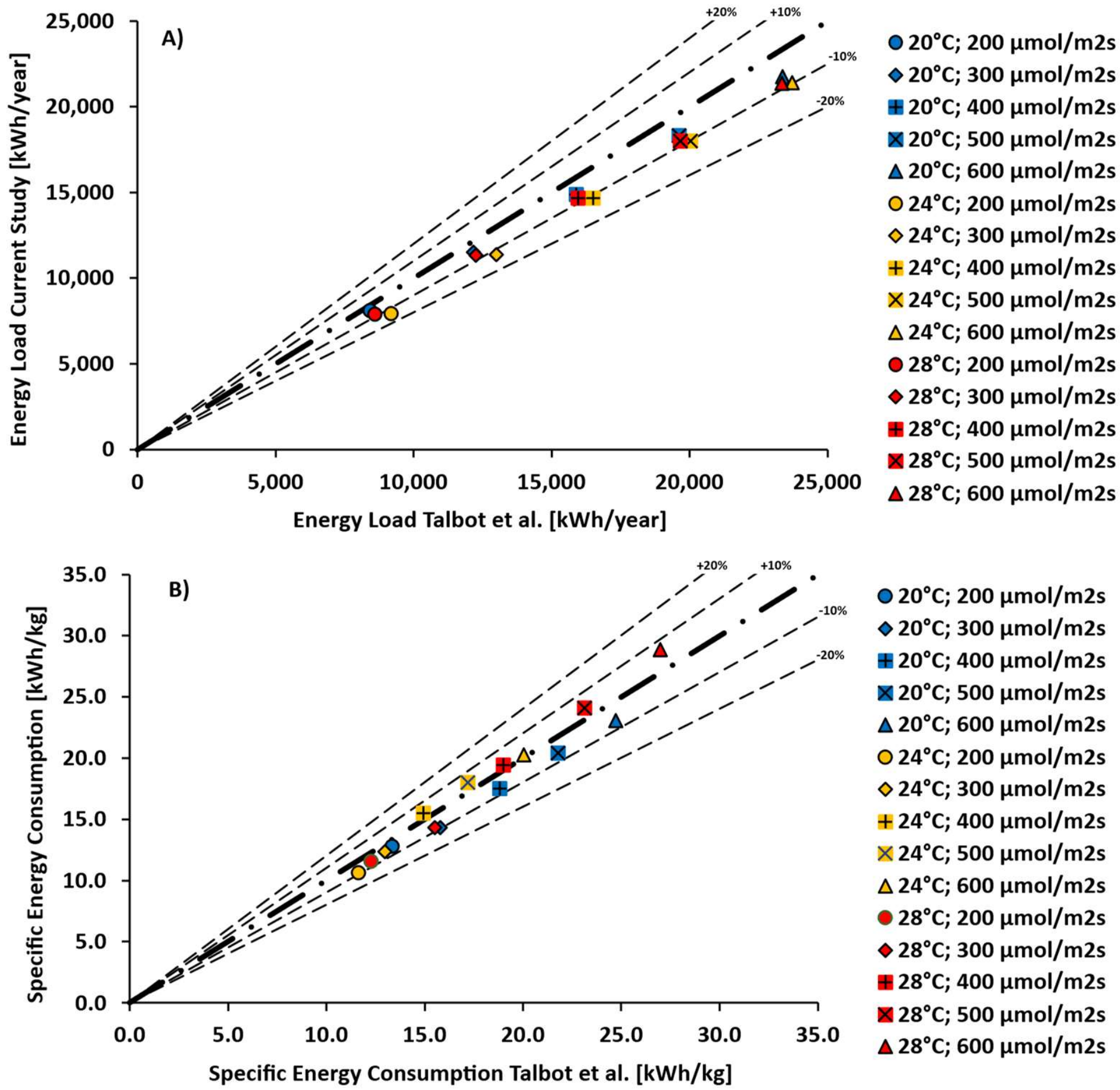

Figure 2: Validation of the AMESim vertical farm model through intermodel comparison with the dynamic model developed by Talbot et al.: annual energy load (a) and specific energy consumption (b). Dashed lines represent ±10% and ±20% deviation ranges.

### 2.2.2. COP Correlations

As described above, the thermal loads of the vertical farm were supplied by an electrically driven HVAC system composed of a heat pump and a chiller. The heat pump was used to cover space heating and post heating requirements, whereas the chiller was used to meet space cooling and dehumidification loads. To estimate the total electricity consumption of the vertical farm, the thermal loads computed by the agri-energy model were converted into electrical loads through performance correlations accounting for outdoor air temperature and part load operation.

For the heat pump, the effect of operating temperatures was represented using a second-law efficiency approach. The nominal second-law efficiency was corrected at each timestep as a function of the ratio between nominal and actual temperature lifts as reported in Eq. (6) [27]. The correction accounts for the degradation of heat pump performance when the temperature lift between the hot water supply and the outdoor air temperature increases. The corrected second law efficiency was then used to calculate the timestep dependent COP according to in Eq. (7), where $T_{hw}$ is the hot water supply temperature and $T_{ext}$ is the outdoor air temperature. Part load operation was included through the heating part load factor ($PLF_H$) [28], evaluated from the ratio between the instantaneous heating demand and the maximum heating capacity available at the same outdoor condition. The latter was estimated using Eq. (8), where $Q_{H,nom}$ is the nominal heating capacity. This formulation allowed the model to account simultaneously for outdoor temperature effects, variable heating capacity and off design operation.

The chiller electricity consumption was evaluated using an analogous approach, although with a simplified treatment of temperature effects. Since the simulated period refers to winter operation, the outdoor air temperature remained below the lower bound of the compressor performance map. Under these conditions, variations in outdoor temperature were not considered to affect the EER, and chiller performance was assumed to depend only on part load operation. The part load correction for the chiller was therefore applied using the relationship reported in [29]. This allowed the cooling and dehumidification loads to be converted into electrical consumption while accounting for off design operation. Further details on the adopted correlations and modeling assumptions are reported in Section 6.4 of the Appendix.

Auxiliary electricity consumption, including pumps and fans associated with the HVAC equipment, was assumed to be included in the nominal COP and EER values. This assumption was considered acceptable because auxiliary loads are expected to represent only a minor fraction of the overall electricity consumption of the vertical farm, which is mainly driven by artificial lighting and climate control.

$$\eta_{II}(t) = \eta_{II,nom} \cdot \left(\frac{\frac{T_{hw,nom}+273.15}{T_{ext,nom}+273.15}}{\frac{T_{hw}+273.15}{T_{ext}(t)+273.15}}\right)^3 \quad (6)$$

$$COP(t) = \eta_{II}(t) \cdot \frac{T_{hw}+273.15}{T_{hw}-T_{ext}(t)} \cdot PLF_H(t) \quad (7)$$

$$Q_{H,max}(t) = Q_{H,nom} \cdot \left(\frac{\frac{T_{hw,nom}+273.15}{T_{ext,nom}+273.15}}{\frac{T_{hw}+273.15}{T_{ext}(t)+273.15}}\right)^3 \quad (8)$$

### 2.3 Data Collection and Conditioning

This section describes the data collection and conditioning procedure adopted to derive a surrogate model suitable for mathematical programming optimization. In this study, the procedure was applied to crop–energy response data generated by the verified AMESim agri-energy model. The original model includes nonlinear relationships related to crop growth, transpiration, HVAC operation, and lighting-dependent energy demand. Although this formulation provides a detailed representation of vertical farm behavior, its direct integration into an optimization framework would lead to a nonconvex problem with high computational complexity.

To overcome this limitation, the main agronomic and energy dependencies were derived and approximated through piecewise linear relationships derived from the AMESim simulation outputs. The surrogate model was designed to preserve the dominant effects of PPFD, crop developmental stage, and outdoor air temperature while reducing the number of variables and constraints required by the optimizer. Different temporal aggregation levels were also considered to evaluate the trade-off between the temporal resolution of the lighting schedule and the computational burden of the mathematical programming formulation.

#### 2.3.1 AMESim Output

To limit the complexity of the surrogate formulation, only the most relevant AMESim outputs were extracted. The selected variables were chosen to retain the main couplings between crop development, lighting operation, and HVAC energy demand.

On the energy side, three output categories were considered. Heating demand was defined as the sum of space-heating and post-heating requirements, while cooling demand included both space-cooling and dehumidification loads. These quantities represent the thermal loads to be supplied by the HVAC system and were later converted into electrical consumption using performance correlations for the heat pump and chiller. Lighting electricity consumption was considered separately, since it depends directly on the imposed PPFD level and on the adopted LED efficiency.

On the agronomic side, the hourly variation of fresh biomass and the hourly variation of leaf area were extracted. Hourly increments, rather than absolute biomass and leaf area values, were used to avoid an explicit dependence on the full previous growth trajectory. In this way, crop growth during a given timestep depends on the PPFD applied during that timestep and on the LAI at the beginning of the timestep, which acts as the state variable summarizing previous canopy development. This allows the optimizer to reconstruct crop development dynamically while preserving the dependence of growth on both light intensity and canopy state.

The energy outputs were parameterized according to their physical drivers. Lighting electricity demand was modeled as a function of PPFD only. Conversely, heating and cooling demands depend on PPFD, LAI, and outdoor air temperature. PPFD affects the heat released by the lighting system and the energy intercepted by the canopy, LAI influences transpiration and latent loads, and outdoor temperature affects heat exchange with the exterior and HVAC performance. Particular attention was given to thermal loads. In the physical AMESim model, heating and cooling powers are affected by the internal control logic of the vertical farm and by the dynamic response of the room and HVAC system. As a result, instantaneous thermal power may vary even when PPFD, LAI, and outdoor temperature remain unchanged. To obtain values suitable for MILP optimization while preserving the overall energy balance of the physical model, heating and cooling powers were averaged over 8 h intervals. This reduced short-term fluctuations caused by control actions and thermal inertia without altering the total energy consumption predicted by AMESim.

#### 2.3.2 Data Conditioning

The AMESim outputs were processed in three steps to represent the time-dependent variability arising from changes in the input values at each time step, while maintaining a linear formulation of the optimization model: interpolation over PPFD values, discretization over LAI values, and temporal aggregation.First, the original AMESim simulations, performed at five PPFD levels equal to 100, 200, 300, 400, and 500 μmol $m^{-2}$ $s^{-1}$, were linearly interpolated to generate additional intermediate light levels. This allowed the surrogate model to describe crop growth and energy consumption also at 50, 150, 250, 350, and 450 μmol $m^{-2}$ $s^{-1}$, increasing the resolution of the lighting decision space without requiring additional physical simulations.

Second, the continuous LAI domain was discretized into 12 representative values. These values define the crop developmental states used by the surrogate model to estimate hourly growth increments and HVAC loads. The number of LAI breakpoints was selected through an iterative comparison between the surrogate model and the original physical AMESim model. Increasing the number of LAI points improved the agreement with the nonlinear formulation but also increased the size of the MILP problem. The adopted discretization was therefore selected as a compromise between approximation accuracy and computational efficiency.

Third, temporal aggregation was applied to reduce the number of optimization timesteps. Starting from the hourly surrogate relationships, 4 h operating blocks were generated for each PPFD level and each initial LAI value of the discretized grid. Within each 4 h block, PPFD was assumed constant. However, LAI was not kept constant. Starting from the initial LAI value, canopy development was updated hour by hour using the hourly LAI increment predicted by the surrogate relationship for the selected PPFD and the LAI reached at the previous hour, as shown in Eq. (9). This allowed the 4 h aggregation to account for crop development occurring within the block.

The reconstructed hourly LAI evolution was then used to compute the aggregated 4 h variation of LAI and fresh biomass. Specifically, $\Delta LAI_{4h}$ and $\Delta FM_{4h}$ were obtained by summing the corresponding hourly increments over the block, as described in Eqs. (10) and (11). Heating and cooling electrical powers were aggregated in the same way. $P_{H,1h}$ and $P_{C,1h}$ were obtained by coupling the thermal loads predicted by the surrogate model with the timestep dependent COP and EER values of the HVAC system. Their 4 h aggregated values, $E_{H,4h}$ and $E_{C,4h}$ were then calculated according to Eqs. (12) and (13). In Eqs. (9) to (13), $\Delta LAI_{1h}$, $\Delta FM_{1h}$, $E_{H,1h}$, and $E_{C,1h}$ are piecewise linear functions derived from the interpolated and discretized AMESim outputs. The growth increments depend on PPFD and current LAI, while heating and cooling powers also depend on outdoor air temperature.

$$LAI(t) = LAI(t-1) + \Delta LAI_{1h}(PPFD, LAI(t-1)) \; \forall t \in \{1,2,3,4\} \quad (9)$$

$$\Delta LAI_{4h} = \sum_{t=1}^{4} \Delta LAI_{1h}(PPFD,\; LAI(t-1)) \quad (10)$$

$$\Delta FM_{4h} = \sum_{t=1}^{4} \Delta FM_{1h}(PPFD,\; LAI(t-1)) \quad (11)$$

$$E_{H,4h} = \sum_{t=1}^{4} E_{H,1h}(PPFD,\; LAI(t-1), T_{ext}) \quad (12)$$

$$E_{C,4h} = \sum_{t=1}^{4} E_{C,1h}(PPFD,\; LAI(t-1), T_{ext}) \quad (13)$$

In addition to the 4 h aggregation, 24 h blocks were generated to describe daily crop development under the reference photoperiod. In this aggregation level, the photoperiod was not treated as an optimization variable. Instead, each daily block was constructed assuming a fixed 16-h light period and an 8-h dark period, consistently with the AMESim reference simulations. The optimizer can therefore select the PPFD applied during the light period of each day, but it cannot modify the intra-day distribution of light hours. As for the 4 h blocks, LAI was updated recursively within the day, so that the resulting daily relationships accounted for canopy development throughout the complete light–dark cycle.

Outdoor-temperature values were selected directly from the temperature profile adopted for the case study, in order to avoid introducing an additional discretization dimension and to limit the computational burden of the surrogate formulation.

Through this procedure, the relevant AMESim outputs and model relationships were first selected to retain the main agronomic and energy-related dependencies of the system. These outputs were then transformed into a set of piecewise linear relationships defined over the discretized LAI grid and different PPFD levels. Depending on the selected aggregation level, namely 1 h, 4 h, or 24 h, the surrogate model maps the PPFD chosen by the optimizer, together with the current LAI and outdoor air temperature, into the corresponding leaf area increase, biomass growth, and electrical demand for lighting, heating, and cooling. Longer aggregation timesteps reduce the temporal resolution of the lighting schedule and as a consequence also decrease the computational burden of the optimization model. Thus, the surrogate model provides a compact and optimization-ready representation of the original agri-energy model, preserving the main interactions between crop growth, lighting operation, and HVAC energy demand.

### 2.4 Problem Statement

The scheduling optimization problem considered in this study determines the optimized lighting schedule over a finite planning horizon T. The problem is defined by a surrogate model describing the crop–energy interactions within the vertical farming system, time-dependent outdoor temperature, and time-dependent electricity prices. For each time period t, the optimization model selects both the on/off status of the lighting system and the corresponding PPFD level. The objective is to minimize the total electricity cost incurred over the full cultivation cycle, subject to the requirement that the required crop production target is met at the end of the planning horizon.

### 2.5 Mathematical Programming Framework

The mathematical programming framework was developed to identify the lighting strategy that minimizes the total electricity cost over the cultivation cycle while ensuring that the crop reaches the target harvest weight. At each optimization timestep, the model selects one PPFD level among the available lighting setpoints. This decision determines the light regime applied to the crop and, through the surrogate formulation, defines the corresponding biomass growth, LAI evolution, lighting electricity consumption, and HVAC energy demand.

The effect of each lighting decision on crop growth and HVAC energy demand is evaluated through the surrogate model derived from the AMESim simulations as piecewise linear correlations. For the selected PPFD level, the model uses the current LAI to interpolate the corresponding fresh biomass increment, LAI increment, heating electricity demand, and cooling electricity demand over the discretized LAI grid. The interpolation is performed using SOS2 variables [30], while binary variables ensure that only one PPFD level is active at each timestep.

The optimization is therefore recursive over the cultivation cycle. Starting from the initial crop state, the selected PPFD determines the biomass and LAI increments at the current timestep. The updated LAI and biomass are then passed to the following timestep, where the same procedure is repeated. In this way, the MILP accounts for the interaction between lighting decisions, canopy development, HVAC loads, electricity prices, and final crop weight.

The same formulation can be applied using different temporal aggregation levels, namely 1 h, 4 h, or 24 h. Shorter timesteps allow a more detailed lighting schedule and a finer representation of crop development, whereas longer timesteps reduce the number of variables and constraints, making the MILP computationally lighter. This allows the tradeoff between optimization resolution and computational effort to be explicitly assessed. It should be noted that the controllable lighting variables differ among aggregation levels. With 1 h and 4 h timesteps, both PPFD and photoperiod can be adjusted by selecting lighting-on or lighting-off periods within each day. Conversely, with 24 h aggregation, the photoperiod is fixed at 16 h $d^{-1}$ and only the PPFD applied during the daily light period is optimized.

In addition to the selection of lighting intensity, two daily agronomic constraints were imposed to ensure that the optimized lighting schedules remained consistent with lettuce cultivation requirements. First, the daily light integral was constrained not to exceed a prescribed maximum value, selected according to values commonly reported for lettuce cultivation under controlled environments [16]. This constraint was introduced because excessively high DLI levels may negatively affect final product quality [13]. This prevents the optimizer from concentrating excessive light within a single day when electricity prices are low. Second, the maximum number of light hours per day was limited to remain within an admissible photoperiod range for lettuce growth. These constraints were imposed at the daily level by grouping all optimization timesteps belonging to the same day.

The MILP formulation is defined over the following sets and indices reported in Table *1*.

Table 1: Definition of sets, indices, and input parameters adopted in the MILP formulation, including lighting levels, crop-state breakpoints, agronomic constraints, and energy-related inputs.

| **Symbol** | **Description** | **Unit** |
|---|---|---|
| $t \in T = \{1, \dots, N\}$ | Optimization timestep within the cultivation cycle, defined according to the selected temporal aggregation level (1 h, 4 h, or 24 h) | [-] |
| $k \in K = \{1, \dots, K\}$ | Discrete PPFD level available to the optimizer as a possible lighting setpoint | [-] |
| $j \in J = \{1, \dots, 12\}$ | Breakpoint of the discretized LAI grid used for piecewise linear interpolation of the surrogate relationships | [-] |
| $d \in D$ | Days of the cultivation cycle | [-] |
| $T_d \subseteq T$ | Set of optimization timesteps belonging to day d | [-] |
| $K^{on} \subseteq K$ | Set of PPFD levels corresponding to lighting-on conditions $(\widehat{p_k} > 0)$ | [-] |
| $\widehat{p_k}$ | PPFD associated with lighting level $k$ | [-] |
| $\widehat{c_t}$ | Average electricity price during timestep $t$ | [€ $MWh^{-1}$] |
| $\widehat{A_{crop}}$ | Crop Area | [$m^2$] |
| $\widehat{T_t}$ | Average outdoor air temperature during timestep $t$ | [°C] |
| $\widehat{\rho}$ | Plant density | [plants $m^{-2}$] |
| $\widehat{W^{target}}$ | Target fresh weight per plant | [g $plant^{-1}$] |
| $\widehat{x_{t,k,J}}$ | LAI breakpoint of the surrogate model for timestep $t$, PPFD level $k$, and discretized LAI point $j$ | [-] |
| $\widehat{f_{t,k,J}^{FM}}$ | Fresh biomass increment derived from the surrogate model | [g $m^{-2}$] |
| $\widehat{f_{t,k,J}^{LAI}}$ | LAI increment derived from the surrogate model | [$m^{-2}$] |
| $\widehat{f_{t,k,J}^{heat}}$ | Heating electrical demand derived from the surrogate model | [Wh $m^{-2}$] |

| $\widehat{f_{t,k,j}^{cool}}$ | Cooling electrical demand derived from the surrogate model | [Wh m$^{-2}$] |
|---|---|---|
| $\widehat{L_k}$ | Lighting electrical demand associated with PPFD level $k$ | [Wh m$^{-2}$] |
| $\widehat{DLI^{max}}$ | Maximum daily lighting integral allowed per day | [mol m$^{-2}$ d$^{-1}$] |
| $H^{\widehat{light,max}}$ | Maximum number of light hours allowed per day | [h d$^{-1}$] |
| $\Delta t$ | Duration of one optimization timestep | [h] |
| $\widehat{dli_k}$ | Light integral supplied during one timestep when PPFD level k is selected | [mol m$^{-2}$] |

The decision variables of the MILP model are summarized in Table *2*. They include continuous variables describing crop development, energy consumption, and operating costs, binary variables used to select the active PPFD level at each timestep, and SOS2 variables used to perform piecewise linear interpolation over the discretized LAI grid.

Table 2: Definition of the continuous, binary, and SOS2 decision variables adopted in the MILP formulation.

| **Symbol** | **Description** | **Unit** |
|---|---|---|
| $LAI_t$ | Leaf area index at the end of timestep t | [-] |
| $FM_t$ | Fresh biomass accumulated at the end of timestep t | [kg m$^{-2}$] |
| $W_t$ | Fresh weight per plant at the end of timestep t | [g plant$^{-1}$] |
| $E_t$ | Total electricity demand during timestep t | [kWh] |
| $C_t$ | Electricity cost during timestep t | [€] |
| $E_t^{cum}$ | Cumulative electricity demand up to timestep t | [kWh] |
| $C_t^{cum}$ | Cumulative electricity cost up to timestep t | [€] |
| $PPFD_t$ | PPFD selected by the optimizer during timestep t | [µmol m$^{-2}$ s$^{-1}$] |
| $E_t^{heat}$ | Heating electricity demand during timestep t | [kWh] |
| $E_t^{cool}$ | Cooling electricity demand during timestep t | [kWh] |
| $L_t$ | Lighting electricity demand during timestep t | [kWh] |
| $z_{t,k} \in \{0,1\}$ | Binary variable equal to 1 if PPFD level k is selected at timestep t | [-] |
| $\lambda_{t,k,j} \geq 0$ | SOS2 interpolation weight associated with LAI breakpoint j and PPFD level k at timestep t | [-] |

The objective of the MILP formulation is to minimize the cumulative electricity cost over the cultivation cycle. The total cost is computed recursively from the timestep electricity demand and the corresponding electricity price signal. The optimization objective is therefore expressed as described in Eq. (14), where $C^{cum}_N$ is the cumulative electricity cost at the final timestep N.

$$f_{obj} = min C_N^{cum} \tag{14}$$

The MILP formulation is constrained to ensure consistency between lighting decisions, daily lighting limits, surrogate model interpolation, crop growth dynamics, and energy cost calculation. At each timestep, the optimizer must select one and only one PPFD level among the available lighting setpoints. This condition is imposed through the binary variables $z_{t,k}$, which activate the selected lighting level and allow the corresponding $PPFD_t$ to be reconstructed, as described in Eqs. (15) and (16). Daily lighting constraints are then imposed to limit both the maximum daily light integral and the maximum number of light hours, as shown in Eqs. (17) and (18). The timestep contribution to DLI is computed from the selected PPFD level and timestep duration, while the photoperiod constraint counts only timesteps in which a positive PPFD level is selected.

The crop response and HVAC electricity demand associated with the selected PPFD are evaluated through piecewise linear interpolation over the discretized LAI grid. The interpolation weights $\lambda_{t,k,j}$ are linked to the active PPFD level and constrained as SOS2 variables, so that interpolation occurs only between adjacent LAI breakpoints, as shown in Eq (19). The current crop state, represented by $LAI_{t-1}$, is then mapped onto the surrogate grid before calculating growth and energy outputs, according to Eq. (20).

Fresh biomass and LAI increments are obtained from the surrogate relationships and used to update crop development recursively over the cultivation cycle, as reported in Eqs. (21)–(24). Similarly, heating and cooling electricity demands are obtained from the surrogate model, while lighting electricity demand depends only on the selected PPFD level, as shown in Eqs. (25)–(27). Total electricity demand at each timestep is then computed as the sum of lighting, heating, and cooling contributions, according to Eq. (28). Electricity cost is calculated by multiplying the timestep electricity demand by the corresponding electricity price, as expressed in Eq. (29). Both energy consumption and cost are accumulated over the optimization horizon using Eqs. (30) and (31). Finally, the harvest constraint ensures that the crop reaches the target fresh weight per plant at the final timestep, as shown in Eqs. (32) and (33).

$$\sum_{k\in K} z_{t,k} = 1 \;\forall t \tag{15}$$

$$PPFD_t = \sum_k \widehat{p_k} \cdot z_{t,k} \tag{16}$$

$$\sum_{t\in T_d} \sum_{k\in K} \widehat{dli}_k \cdot z_{t,k} \le D\widehat{LI^{max}} \;\forall d \in D \tag{17}$$

$$\sum_{t\in T_d} \sum_{k\in K^{on}} \Delta t \cdot z_{t,k} \le H^{\widehat{light,max}} \;\forall d \in D \tag{18}$$

$$\sum_j \lambda_{t,k,j} = z_{t,k},\; \lambda_{t,k,\cdot} \in SOS2 \tag{19}$$

$$LAI_{t-1} = \sum_k \sum_j \widehat{x_{t,k,j}} \cdot \lambda_{t,k,j} \tag{20}$$

$$\Delta FM_t = \sum_k \sum_j \widehat{f_{t,k,j}^{FM}} \cdot \lambda_{t,k,j} \tag{21}$$

$$\Delta LAI_t = \sum_k \sum_j \widehat{f_{t,k,j}^{LAI}} \cdot \lambda_{t,k,j} \tag{22}$$

$$FM_t = FM_{t-1} + \Delta FM_t \tag{23}$$

$$LAI_t = LAI_{t-1} + \Delta LAI_t \tag{24}$$

$$E_t^{heat} = \frac{\widehat{A_{crop}}}{1000} \sum_k \sum_j \widehat{f_{t,k,j}^{heat}} \cdot \lambda_{t,k,j} \tag{25}$$

$$E_t^{cool} = \frac{\widehat{A_{crop}}}{1000} \sum_k \sum_j \widehat{f_{t,k,j}^{cool}} \cdot \lambda_{t,k,j} \tag{26}$$

$$L_t = \frac{\widehat{A_{crop}}}{1000} \sum_k \widehat{L_k} \cdot z_{t,k} \tag{27}$$

$$E_t = E_t^{he} \quad + E_t^{cool} + L_t \tag{28}$$

$$C_t = \widehat{c_t} \cdot \frac{E_t}{1000} \tag{29}$$

$$E_t^{cum} = E_{t-1}^{cum} + E_t \tag{30}$$

$$C_t^{cum} = C_{t-1}^{cum} + C_t \tag{31}$$

$$W_t = \frac{1000 \cdot FM_t}{\widehat{\rho}} \tag{32}$$

$$W_N \geq \widehat{W^{target}} \quad (33)$$

### 2.6 Case Study

The surrogate model derived from the synthetic data generated by the physical AMESim model was used in this study to identify optimized lighting schedules aimed at minimizing the electricity cost of the vertical farm while ensuring that the crop reached the target harvest weight. To assess the impact of optimized scheduling, two benchmark lighting scenarios were selected. Both scenarios were based on a fixed photoperiod of 16 h $d^{-1}$ and different constant PPFD levels, as reported in Table *3*, resulting in crop cycle durations of 26 and 21 days, respectively.

Table 3: Definition of the fixed-lighting benchmark scenarios used to evaluate the optimized lighting schedules.

| Scenario | PPFD Fix [μmol $m^{-2}$ $s^{-1}$] | Photoperiod [h] | Crop Cycle [day] |
|---|---|---|---|
| Bench_200 | 200 | 16 | 26 |
| Bench_300 | 300 | 16 | 21 |

For each benchmark scenario, the optimization framework was applied over the same crop cycle duration, allowing the PPFD to be adjusted at each optimization timestep. Three temporal aggregation levels were considered, namely 24 h, 4 h, and 1 h, in order to evaluate the effect of scheduling resolution on the achievable cost reduction. At each timestep, the optimized PPFD was selected as a function of electricity price, outdoor temperature, and crop developmental state, represented by LAI. For the 1 h and 4 h formulations, the optimizer could vary both PPFD and photoperiod by selecting lighting-on and lighting-off periods within each day. Conversely, in the 24 h formulation, the photoperiod was fixed at 16 h $d^{-1}$ and only the PPFD applied during the daily light period was optimized. Therefore, additional daily agronomic constraints were imposed on the 1 h and 4 h formulations to prevent unrealistic lighting schedules. Specifically, the daily light integral was limited to 17.3 mol $m^{-2}$ $d^{-1}$, while the maximum number of light hours was set to 20 h $d^{-1}$. These constraints were imposed to restrict the solution space to operating conditions broadly representative of commercial vertical farming practice and to avoid potentially unrealistic or agronomically undesirable solutions. This is particularly important because the optimization model captures crop yield and energy consumption but does not explicitly represent the effects of DLI and photoperiod on final product quality.

The case study was located in northern Italy, where large-scale industrial vertical farms are already present. The outdoor temperature profile, reported in Appendix 6.5, was derived from PVGIS data [31]. The electricity price profile, shown in Figure *3*, was based on the day-ahead market price for northern Italy [32] and was adjusted to include fees and additional taxes using the average electricity price for industrial users in Italy provided by [33]. These assumptions were considered suitable for the purpose of this study, since the analysis focuses on the relative performance of optimized lighting schedules compared with conventional benchmark strategies under the same boundary conditions. The size of the optimization problem depends directly on both the length of the planning horizon, equal to 21 or 26 days, and the adopted temporal discretization, equal to 1, 4, or 24 h. Finer time resolutions increase the number of decision periods and therefore replicate state variables, control variables, interpolation variables, and linking constraints across the entire crop cycle. In the largest instance, the formulation includes 60,531 variables, of which 4,368 are binary, and 9,415 constraints. Although this problem size is not exceptionally large in absolute terms, its temporal structure makes it computationally demanding. Crop productivity is cumulative, and the biomass and LAI reached in each period depend on all previous lighting decisions. Moreover, the final production requirement couples the complete sequence of decisions over the planning horizon, since a locally inexpensive decision may reduce crop growth and require more intensive and costly operation in later periods. This produces a tightly coupled dynamic problem with a weakly separable structure and a large number of feasible scheduling combinations generated by the binary variables. As a result, the solver must evaluate decisions not only according to their immediate energy cost but also according to their effect on the remaining production trajectory and on the feasibility of the terminal productivity constraint. These features substantially increase branch and bound effort and solution time. They would be even more problematic in a direct nonlinear formulation, where the same intertemporal coupling would be combined with nonlinear state transitions, possible nonconvexity, poorly conditioned gradients, and multiple stationary points, further limiting convergence reliability and eliminating general guarantees of global optimality.

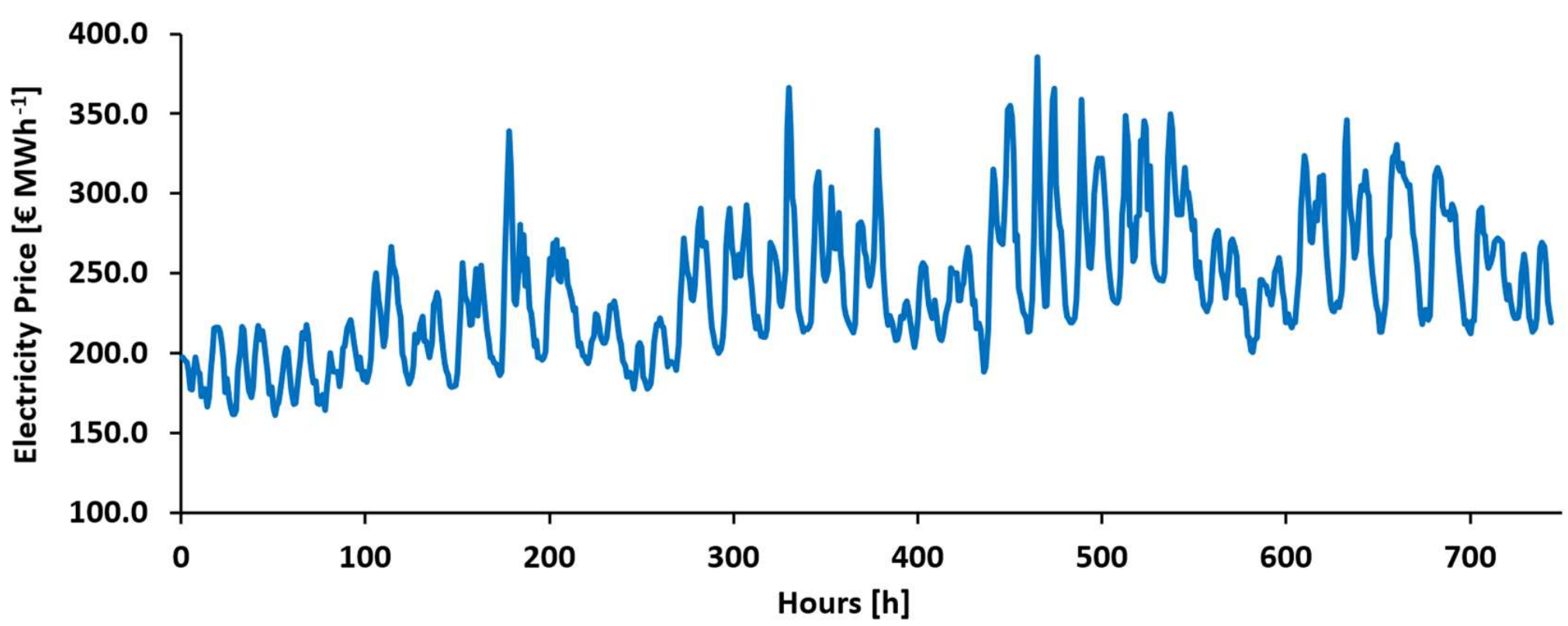

Figure 3: Electricity cost profile adopted for the selected case-study location during a representative winter month and used as input to the optimization framework.

## 3. Results

This section presents the results obtained from the proposed computational procedure for deriving and applying a piecewise linear surrogate model for crop and energy analysis. The analysis is organized into four parts. First, the general procedure used to generate the surrogate model is summarized, highlighting the main steps required to transform crop and energy response data into input–output relationships suitable for optimization. Second, the consistency of the generated surrogate model is evaluated by comparison with the physical AMESim model, which was used in this study as the reference model to generate the original crop and energy response data. The comparison focuses on crop growth and energy demand under different lighting conditions and temporal aggregation levels. Third, the derived piecewise linear relationships are analyzed as compact performance maps, describing the dependence of vertical farm behavior on PPFD, crop developmental stage, and outdoor temperature. Finally, the surrogate formulation is integrated into the MILP framework to assess preliminary optimized lighting schedules against fixed lighting benchmark scenarios.

### 3.1. General procedure for surrogate model generation and optimization

The first outcome of the study is the definition of a general procedure for transforming crop and energy response data into surrogate formulations suitable for optimization. The objective is to provide a workflow that can be extended to different types of vertical farms, supporting the generation of compact surrogate models from either monitored operational data or simulation-based response data. Therefore, the procedure is not restricted to the specific vertical farm configuration analyzed in this study but provides a general approach for translating complex crop and energy interactions into input–output relationships that can be embedded in optimization models. Figure *4* summarizes the proposed workflow, from data acquisition or generation to surrogate construction, consistency assessment, and MILP-based optimization.

In the general framework, the procedure starts from a vertical farm or reference system for which operating inputs, crop outputs, and energy outputs are available. The first step is the identification of the control input to be optimized. This choice is relevant because the simultaneous optimization of multiple control variables can significantly increase the size and complexity of the optimization problem. Therefore, the procedure initially focuses on the variable expected to have the strongest impact on both crop growth and energy demand. In an operating facility, crop and energy data can be collected through a sensor and monitoring system, whereas in simulation-based applications they can be generated by a verified dynamic model.

A second key step is the identification of a growth-state variable. In a temporal scheduling problem, the effect of a control input depends on the physiological state reached by the crop during the growth cycle. Linking the optimization horizon to a crop development indicator allows the surrogate model to account for the fact that the same control action may produce different growth and energy responses at different stages of biomass accumulation. In this study, LAI was selected as the growth-state variable, since it summarizes canopy development and directly influences light interception and transpiration.

Energy outputs can be monitored separately for the main subsystems of the vertical farm. In a real facility, the electricity demand of lighting and HVAC units can be measured using current transformers, power meters, or equivalent monitoring devices. These measurements, together with the crop response data, are then processed to identify the input–output relationships required to construct a compact surrogate formulation suitable for optimization.

Once the growth-state variable has been identified, the temporal profiles of the crop and energy outputs are analyzed to define a suitable number of breakpoints along the crop development trajectory. These breakpoints must be sufficient to describe the full growth process and to capture how the instantaneous crop response and energy demand vary as functions of both the growth state and the selected control input. The resulting set of input–output relationships constitutes the piecewise surrogate model, in which crop growth and energy demand are represented through discrete operating conditions and interpolation over the growth-state grid.

Before being used for optimization, the surrogate model must be checked against the same reference dataset used for its generation, which may consist of sensor measurements from an operating vertical farm or outputs from a validated simulation model. This consistency assessment verifies whether the surrogate formulation can reproduce the main crop and energy responses with sufficient accuracy for optimization purposes. Increasing the number of breakpoints generally improves the accuracy and robustness of the surrogate model, but it also increases the number of interpolation variables and constraints in the MILP formulation. Therefore, the breakpoint grid must be selected as a compromise between approximation accuracy and computational tractability. In the present application, this trade-off led to the adoption of a 12-point LAI grid. To further reduce computational burden, the same breakpoint locations were adopted for all surrogate-model relationships. Using a common grid provides a more regular and compact formulation, simplifies the generation of interpolation and SOS2 constraints, and avoids the additional complexity associated with managing different discretizations for each relationship.

Once the consistency of the piecewise surrogate model has been verified, the resulting formulation can be embedded into a MILP framework to optimize the temporal schedule of the selected control input. In this way, the same variable used to generate the surrogate response relationships is subsequently optimized over the cultivation cycle, providing an operating strategy that accounts for crop development, energy demand, and external boundary conditions.

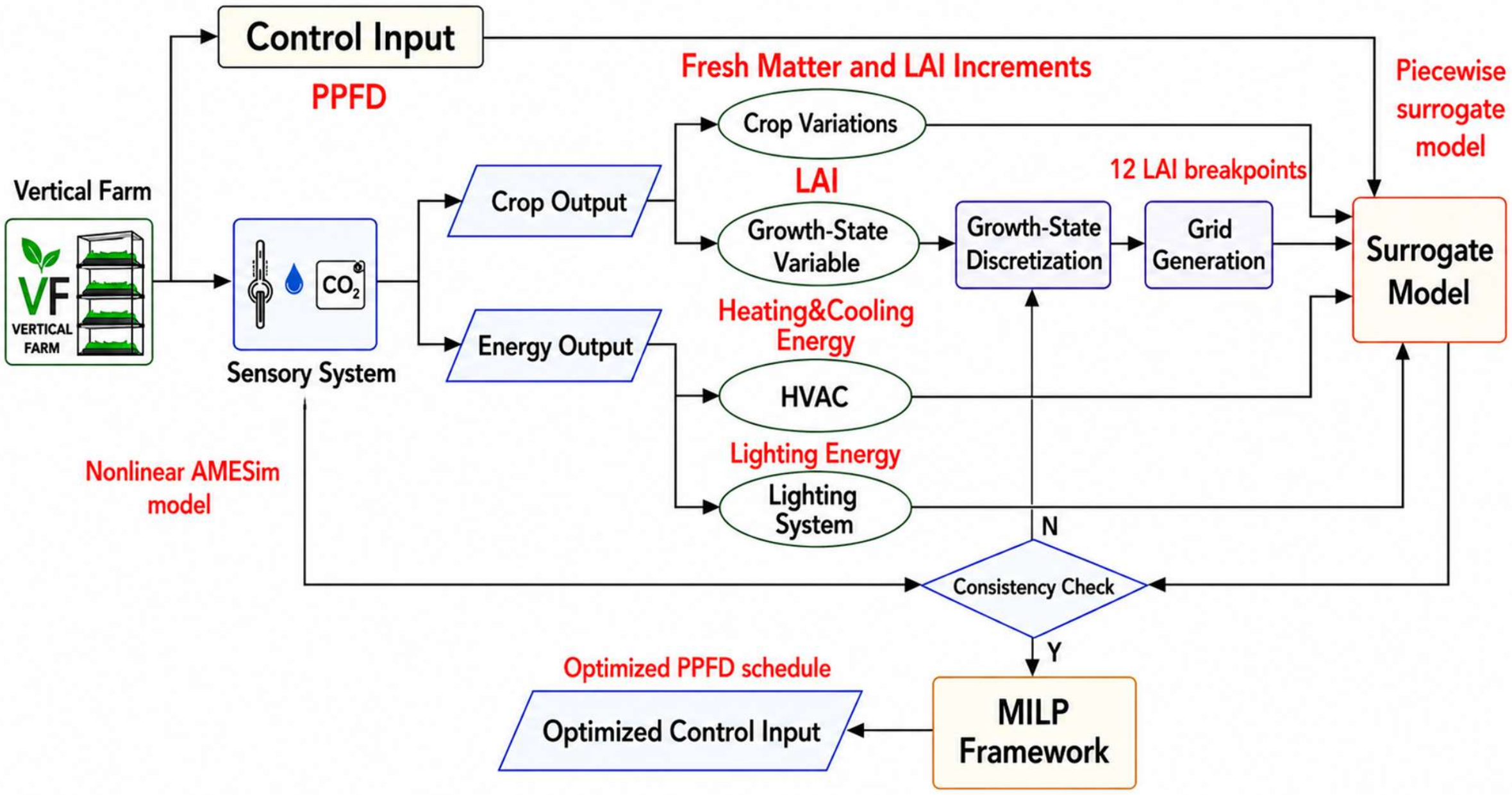

Figure 4: Generalized workflow for converting crop–energy response data into a surrogate model suitable for optimizing a vertical farm control variable. Red labels indicate the corresponding implementation adopted in this study, where synthetic data from the validated physical AMESim model were used instead of an experimental dataset, and PPFD was selected as the optimized control variable.

### 3.2. Surrogate model consistency with the physical reference model

Prior to its integration into the MILP framework, the surrogate formulation was assessed against the physical reference model to quantify its ability to reproduce the relevant crop–energy outputs. The adopted 12-point LAI discretization was selected on the basis of preliminary grid-refinement tests as a trade-off between approximation accuracy and computational tractability. Finer discretizations led to only marginal improvements in surrogate accuracy, while increasing the number of interpolation variables and constraints in the optimization problem.

The consistency assessment was performed under the reference lighting regime, considering a fixed photoperiod of 16 h $d^{-1}$ and five constant PPFD levels, namely 100, 200, 300, 400, and 500 µmol $m^{-2}$ $s^{-1}$. The piecewise linear surrogate model was benchmarked against the physical AMESim model used to generate the original crop–energy response dataset. The purpose of this assessment was to verify that the surrogate formulation retained the dominant crop-growth and energy-demand responses of the physical model with an accuracy suitable for subsequent MILP-based optimization. The models were compared in terms of fresh biomass production, expressed in g $m^{-2}$, and total energy demand, expressed in kWh. Total energy demand included the electricity required for artificial lighting and the thermal energy needed to meet heating and cooling loads. These quantities were selected because they represent the main agronomic and energy-related outputs subsequently used in the MILP framework.

The consistency assessment was also extended to the temporal aggregation procedure, since the surrogate formulation can be implemented with different timestep resolutions in the MILP framework. For conciseness, the results are reported for the two limiting aggregation levels considered in this study, namely the 1 h and 24 h formulations. The former provides the highest temporal resolution and the closest representation of the original AMESim outputs, whereas the latter represents the most compact formulation, in which crop development and energy demand are described at daily scale. Comparing these two cases allows the effect of temporal aggregation on surrogate accuracy to be quantified.

The agreement between the surrogate and physical models was evaluated using multiple statistical indicators. For each tested PPFD level and for both temporal aggregation levels, crop growth and total energy demand were assessed in terms of root mean square error (RMSE), coefficient of determination ($R^2$), and coefficient of variation of the RMSE, CV(RMSE). RMSE quantifies the absolute deviation between surrogate and nonlinear predictions over the crop cycle, whereas $R^2$ measures the ability of the surrogate model to reproduce the variability of the nonlinear reference outputs. CV(RMSE), computed by normalizing the RMSE with respect to the mean value of the nonlinear reference output, provides a relative measure of the approximation error and enables comparison across different PPFD levels.

In addition to the PPFD-specific indicators, global RMSE, $R^2$, and CV(RMSE) values were computed by pooling all timestep predictions obtained across the five tested PPFD levels. These global metrics provide an overall assessment of the surrogate accuracy over the complete crop cycle and across the full lighting range considered in the study. Since fresh biomass and total energy demand are cumulative variables, local deviations in timestep biomass and energy increments may propagate over the crop cycle and affect the final values predicted at harvest. Therefore, an additional harvest-level error was evaluated by computing the RMSE and CV(RMSE) across the final biomass and cumulative energy values obtained for the five tested PPFD levels. This metric directly assesses the accuracy of the surrogate model in predicting the final agronomic and energy outputs used in the subsequent MILP optimization. Since fresh matter is a cumulative state variable, global indicators such as RMSE, $R^2$, and CV(RMSE) may not fully describe the largest local deviation occurring during the crop cycle or its effect on the final harvest prediction. Therefore, the surrogate accuracy was further assessed using the maximum absolute error over the cultivation cycle and the relative error at harvest. The former identifies the largest deviation between the surrogate and physical-model trajectories, while the latter directly quantifies the accuracy of the final fresh-matter prediction used in the optimization framework.

From a crop growth perspective, Figure *5* compares the fresh biomass predicted by the surrogate and physical AMESim models at four representative stages of the crop cycle, corresponding to 25, 50, 75, and 100% of plant growth. The comparison includes all investigated PPFD levels, allowing the surrogate accuracy to be evaluated under different lighting intensities. Overall, the surrogate model reproduced the nonlinear biomass trajectories with good agreement and captured the increase in biomass accumulation associated with higher PPFD values for both temporal aggregation levels. The agreement depended on the selected temporal resolution. The 1 h formulation provided the closest match to the AMESim outputs, while the 24 h formulation showed a slight overestimation of accumulated biomass, particularly toward the end of the crop cycle. This behavior is consistent with the coarser representation of intra-day crop development adopted in the daily-scale formulation.

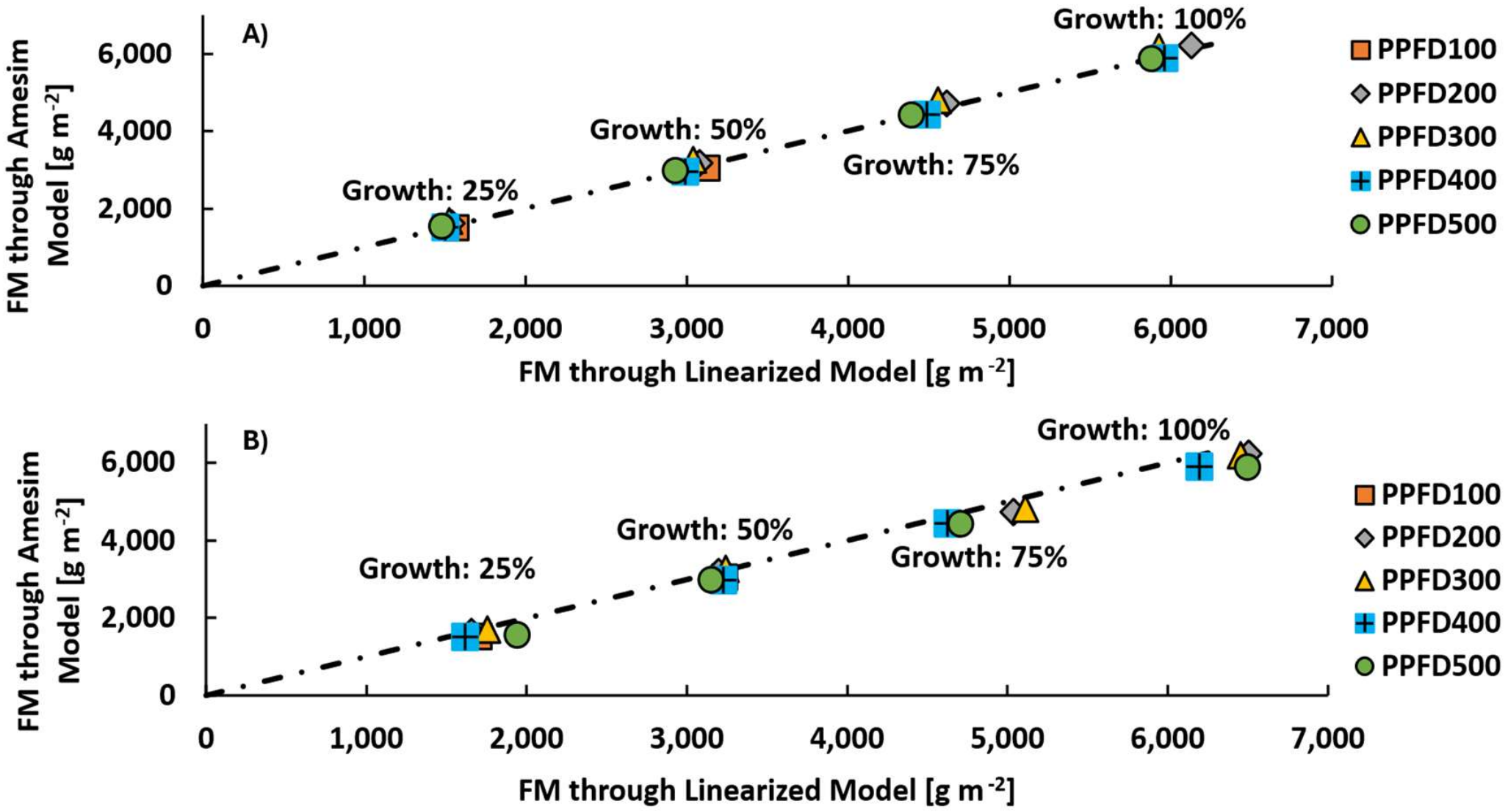

Figure 5: Comparison between fresh biomass predicted by the piecewise linear surrogate model and the physical AMESim model at different crop growth stages and PPFD levels, considering the 1 h (a) and 24 h (b) temporal aggregations.

The effect of temporal aggregation is also reflected in the statistical indicators reported in Table *4*. Both aggregation levels achieved high coefficients of determination, with $R^2$ values above 0.95, confirming that the surrogate model retained the main trends of the nonlinear biomass trajectories. However, reducing the temporal resolution increased the relative approximation error. The global CV(RMSE), computed by pooling all timestep predictions across the five PPFD levels, increased from 1.3% for the 1 h aggregation to 10.8% for the 24 h aggregation.

The surrogate accuracy also varied across PPFD levels. This is partly related to the adoption of a single LAI discretization grid for the entire PPFD range. Although PPFD-specific LAI grids could improve local accuracy, a common discretization was retained to preserve a compact and consistent surrogate structure for MILP implementation.

Since biomass accumulation is cumulative, the error was also evaluated at harvest. Across the five tested PPFD levels, the 1 h aggregation achieved a harvest-level RMSE of 130.4 g $m^{-2}$, corresponding to a CV(RMSE) of 2.3%. The 24 h aggregation showed a higher harvest-level deviation, with an RMSE of 414.0 g $m^{-2}$ and a CV(RMSE) of 7.2%. These results quantify the trade-off between surrogate accuracy and temporal compactness and support the use of the proposed formulation in the subsequent MILP-based optimization.

Table 4: Statistical indicators used to assess the consistency of fresh biomass predictions from the piecewise linear surrogate model against the physical AMESim model for the 1 h and 24 h temporal aggregations.

| **Temporal Aggregation** | | **1 h** | | | **24 h** | |
|---|---|---|---|---|---|---|
| **PPFD Level** | **RMSE** | **R2** | **CV(RMSE)** | **RMSE** | **R2** | **CV(RMSE)** |
| 100 | 42.6 | 0.998 | 4.8% | 93.6 | 0.989 | 11.8% |
| 200 | 74.9 | 0.998 | 3.8% | 67.1 | 0.999 | 3.5% |
| 300 | 147.9 | 0.992 | 9.5% | 75.4 | 0.998 | 5.0% |
| 400 | 29.4 | 0.999 | 1.6% | 238.6 | 0.986 | 14.3% |
| 500 | 35.1 | 0.999 | 2.2% | 246.4 | 0.982 | 16.8% |
| All | 76.9 | 0.998 | 1.3% | 154.6 | 0.992 | 10.8% |

As shown in Figure *6*Maximum absolute fresh-matter error over the cultivation cycle (bars, left axis) and relative fresh-matter error at harvest (markers, right axis) for the 1 h and 24 h surrogate formulations at different PPFD levels. The harvest relative error was calculated from the difference between surrogate- and physical-model fresh matter at the end of the corresponding crop cycle., the additional error indicators confirm the effect of temporal aggregation on surrogate accuracy. The 24 h formulation produced maximum absolute fresh-matter errors ranging from 311.5 to 955.6 g $m^{-2}$, with the largest deviations occurring at 400 and 500 µmol $m^{-2}$ $s^{-1}$. By contrast, the 1 h formulation limited the maximum error to 256.6 g $m^{-2}$ or less across the entire PPFD range. The harvest-level relative error remained below 4.5% for all cases except the 24 h formulation at 400 µmol $m^{-2}$ $s^{-1}$, where it reached 12.95%; for the 1 h formulation, it remained between 0.08% and 4.07%. Overall, these results support the accuracy of the surrogate model, particularly at hourly resolution, while quantifying the larger cumulative deviations introduced by daily aggregation.

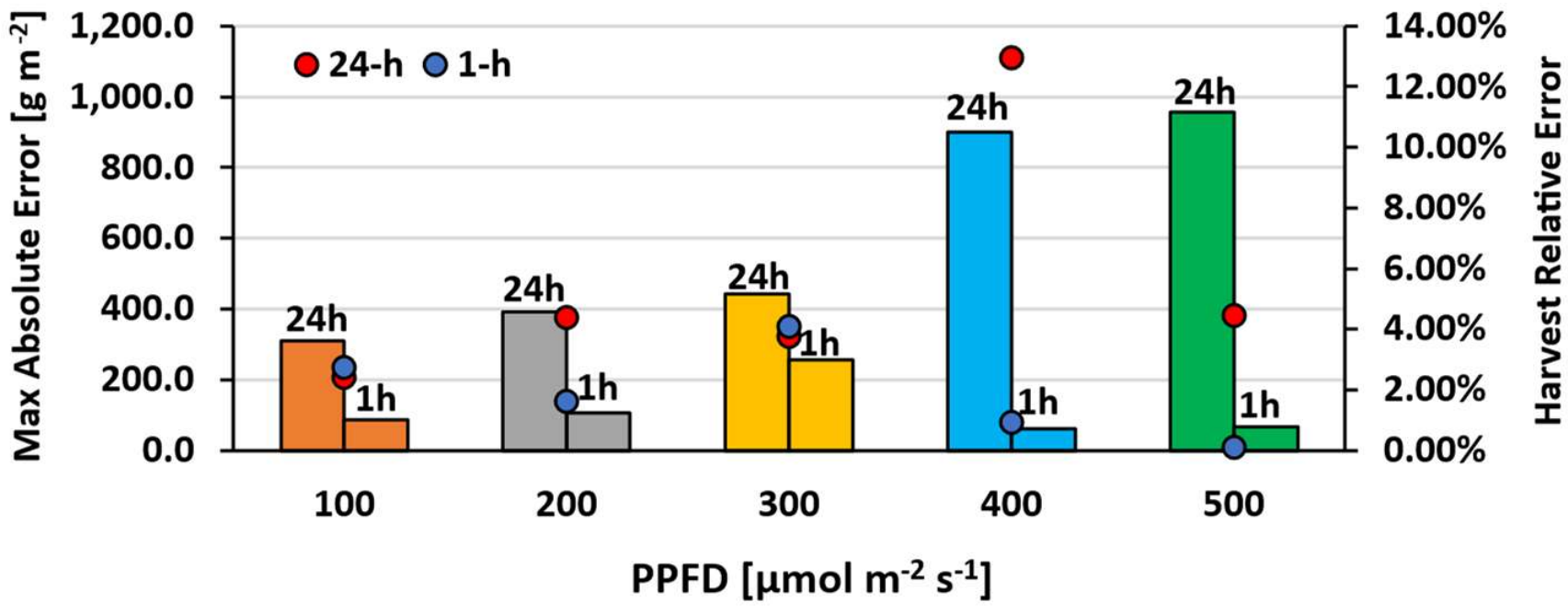

Figure 6: Maximum absolute fresh-matter error over the cultivation cycle (bars, left axis) and relative fresh-matter error at harvest (markers, right axis) for the 1 h and 24 h surrogate formulations at different PPFD levels. The harvest relative error was calculated from the difference between surrogate- and physical-model fresh matter at the end of the corresponding crop cycle.

From an energy perspective, Figure *7* reports the cumulative total energy demand predicted by the surrogate and physical AMESim models at four representative stages of the crop cycle, corresponding to 25, 50, 75, and 100% of plant growth. The comparison includes lighting electricity demand together with the thermal demand required for heating and cooling, before conversion into HVAC electricity consumption. Overall, the surrogate model reproduced the cumulative energy-demand trends with satisfactory agreement across the investigated PPFD range. The agreement was slightly lower than that obtained for fresh biomass, as heating and cooling loads in the physical AMESim model are influenced by short-term thermal dynamics and environmental control actions. These effects were intentionally smoothed in the surrogate formulation by using averaged power values at each breakpoint, in order to derive relationships suitable for MILP integration. Consequently, the surrogate model was designed to preserve the cumulative energy behavior of the physical model rather than reproduce its instantaneous thermal-load fluctuations. As for crop growth, the agreement depended on the temporal aggregation level. The 1 h formulation provided the closest match to the nonlinear outputs, while the 24 h formulation showed a moderate overestimation of cumulative energy demand, especially toward the end of the crop cycle. This deviation is consistent with the daily-scale representation of crop development and energy demand adopted in the 24 h formulation

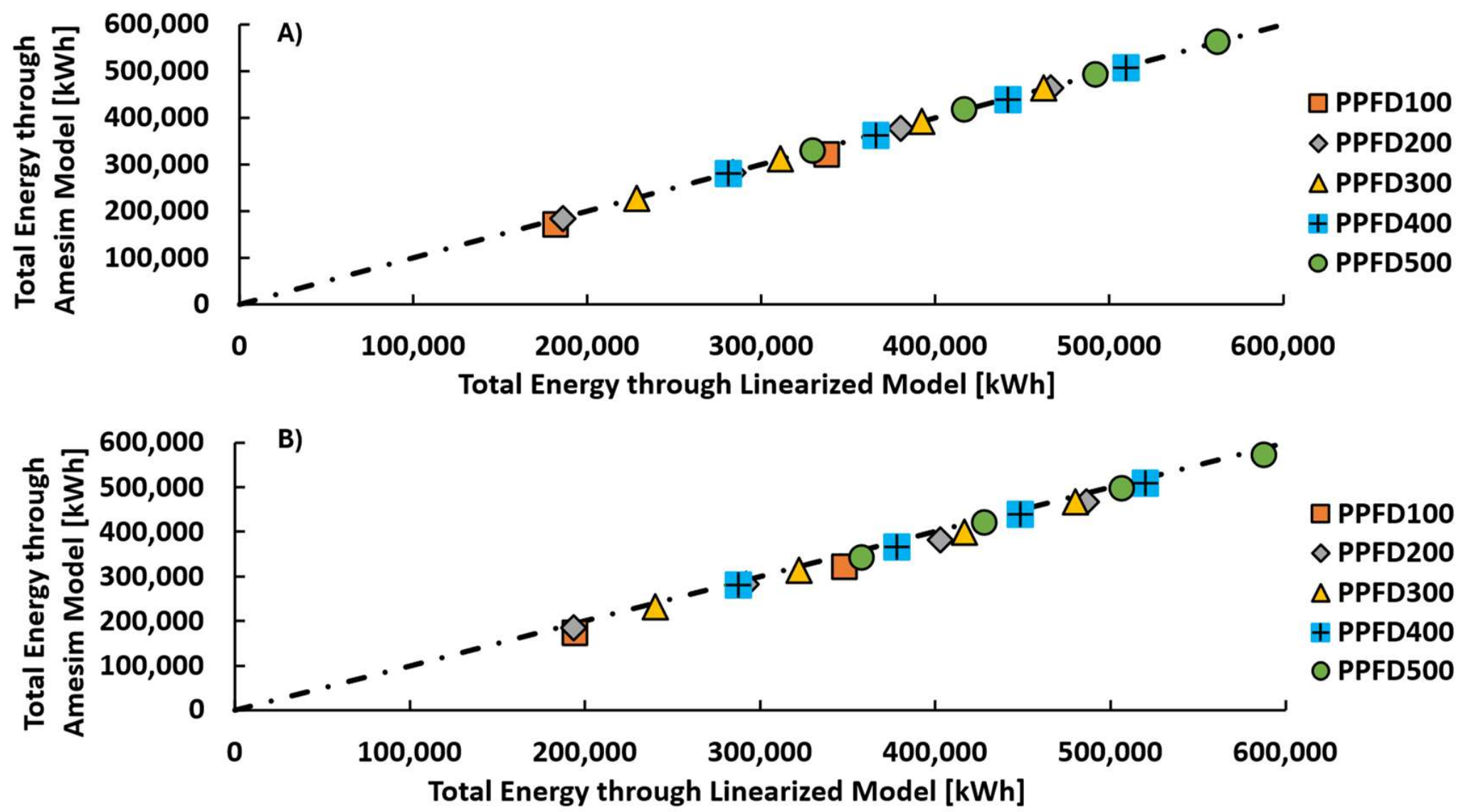

Figure 7: Comparison between total energy demand predicted by the piecewise linear surrogate model and the physical AMESim model at different crop growth stages and PPFD levels, considering the 1 h (a) and 24 h (b) temporal aggregations.

The statistical indicators reported in Table *5* confirm the ability of the surrogate formulation to reproduce the main cumulative energy-demand trends. For both temporal aggregation levels, $R^2$ remained above 0.95, indicating that the surrogate model captured the variability of the physical AMESim outputs over the investigated PPFD range. The relative error was slightly affected by the temporal resolution: the global CV(RMSE), computed by pooling all timestep predictions across the five PPFD levels, increased from 2.5% for the 1 h formulation to 6.3% for the 24 h formulation.

Compared with crop growth, the energy-demand approximation is influenced not only by the LAI discretization, but also by the smoothing procedure adopted for the thermal loads. In the physical AMESim model, heating and cooling demands are affected by short-term room dynamics and environmental control actions. These effects were averaged in the surrogate formulation to obtain stable energy-demand relationships suitable for optimization. As a result, part of the deviation between the two models is associated with the deliberate removal of short-term fluctuations rather than with the inability of the surrogate to reproduce the overall energy balance.

Since total energy demand is accumulated over the crop cycle, the final cycle energy was also evaluated across the five tested PPFD levels. For the 1 h aggregation, the final-cycle RMSE was 8,450 kWh, corresponding to a CV(RMSE) of 1.8%. For the 24 h aggregation, the RMSE increased to 17,167 kWh, with a CV(RMSE) of 3.6%. These values indicate that, despite the simplified representation of thermal dynamics, the surrogate formulation preserved the cumulative energy demand with sufficient accuracy for the subsequent MILP-based scheduling analysis.

Table 5: Statistical indicators used to assess the consistency of total energy demand predictions from the piecewise linear surrogate model against the physical AMESim model for the 1 h and 24 h temporal aggregations.

| **Temporal Aggregation** | **1 h** | | | **24 h** | | |
|---|---|---|---|---|---|---|
| **PPFD Level** | **RMSE** | **R2** | **CV(RMSE)** | **RMSE** | **R2** | **CV(RMSE)** |
| 100 | 8,650 | 0.990 | 7.7% | 12,450 | 0.981 | 11.7% |
| 200 | 2,400 | 0.999 | 1.2% | 7,675 | 0.997 | 4.1% |
| 300 | 840 | 0.999 | 0.4% | 9,600 | 0.994 | 5.3% |
| 400 | 2,460 | 0.999 | 1.0% | 13,000 | 0.994 | 5.0% |
| 500 | 790 | 0.999 | 0.3% | 14,800 | 0.992 | 5.8% |
| All | 4,770 | 0.999 | 2.5% | 11,610 | 0.993 | 6.3% |

The same additional indicators were also evaluated for cumulative energy demand, since global statistics may not fully capture the largest deviation occurring during the cultivation cycle or the accuracy of the final energy estimate. Figure *8* Maximum absolute cumulative-energy error over the cultivation cycle (bars, left axis) and relative cumulative-energy error at harvest (markers, right axis) for the 1 h and 24 h surrogate formulations at different PPFD levels. The harvest relative error was calculated from the difference between surrogate- and physical-model cumulative energy demand at the end of the corresponding crop cycle. reports the maximum absolute error over the crop cycle and the relative error at harvest for the 1 h and 24 h formulations. The daily aggregation produced maximum absolute errors between 22,436 and 36,622 kWh, whereas the hourly formulation remained below 18,426 kWh and below 5,525 kWh for all PPFD levels above 100 µmol $m^{-2}$ $s^{-1}$. The final relative error remained below 4.5% for the 24 h formulation, except at 100 µmol $m^{-2}$ $s^{-1}$, where it reached 5.10%. For the 1 h formulation, it was below 1% for all PPFD levels except 100 µmol $m^{-2}$ $s^{-1}$, where it reached 5.38%. Overall, the results confirm that the surrogate model preserves cumulative energy demand with good accuracy, while the 1 h formulation provides the closest agreement over most of the investigated PPFD range.

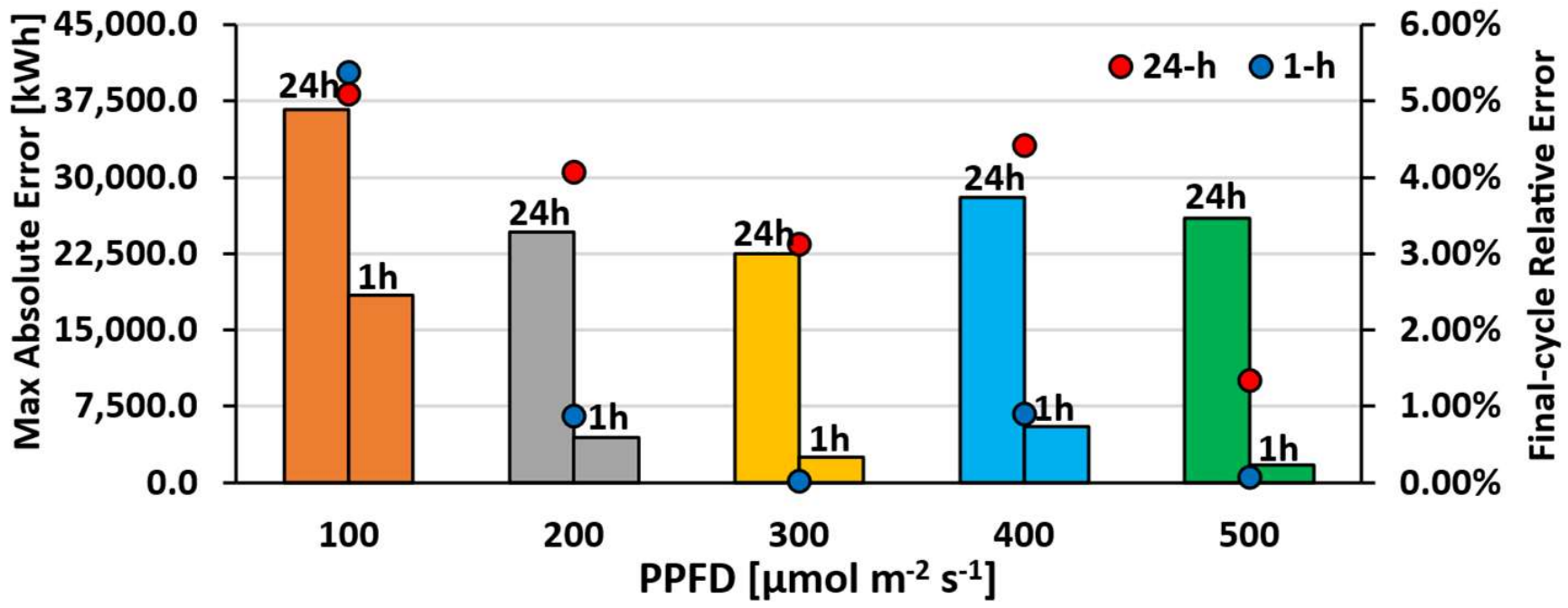

Figure 8: Maximum absolute cumulative-energy error over the cultivation cycle (bars, left axis) and relative cumulative-energy error at harvest (markers, right axis) for the 1 h and 24 h surrogate formulations at different PPFD levels. The harvest relative error was calculated from the difference between surrogate- and physical-model cumulative energy demand at the end of the corresponding crop cycle.

### 3.3. Surrogate model: growth and energy demand piecewise maps

This section presents the piecewise relationships used to describe crop growth and energy demand at each optimization timestep as functions of the selected PPFD level and the current LAI. For brevity, only the maps obtained for the 1 h temporal aggregation are reported in the main text, since this formulation provides the highest temporal resolution and the closest representation of the original AMESim outputs. Figure *9* shows the piecewise maps of fresh biomass increment, $(\widehat{f_{t,k,j}^{FM}})$ and LAI increment $(\widehat{f_{t,k,j}^{LAI}})$ as functions of the discretized LAI grid and PPFD level. The response maps can be interpreted as discrete samples of the continuous nonlinear relationships generated by the physical model. A direct integration of these relationships into the optimization framework would therefore require fitting continuous response surfaces, such as multivariate polynomial functions, to describe crop growth and heating and cooling electricity demand as functions of LAI and PPFD. The observed saturation, threshold effects, extended plateau regions, and changes in slope across the operating domain imply that such fitted surfaces would be highly nonlinear and could not generally be assumed to be jointly convex or concave. In particular, the heating demand relationship exhibits marked changes in curvature and regions in which the demand becomes zero, whereas the crop growth relationships show progressively decreasing marginal responses as LAI increases. High order polynomial fits capable of reproducing these patterns may introduce inflection points, multiple stationary points, oscillatory behavior between sampled values, and poorly conditioned derivatives. When embedded as equality constraints in the multiperiod crop state and energy balance equations, these functions would generate a nonlinear and potentially nonconvex feasible region. The resulting nonlinear programming problem could therefore be sensitive to initialization, converge to locally optimal stationary points, and provide no general guarantee of global optimality. Furthermore, flat regions and steep transitions may lead to weak or ill conditioned gradients, thereby slowing convergence and increasing numerical instability. For these reasons, the original nonlinear response surfaces were replaced by piecewise linear surrogate relationships, enabling their integration into a more tractable optimization formulation.

A common LAI grid was used for all PPFD levels in order to preserve a compact and consistent surrogate structure within the MILP framework. This choice avoids the need to define PPFD-specific interpolation grids, which would increase the complexity of the optimization model and make the surrogate formulation less uniform. The adopted grid is denser at low LAI values, where the crop response changes more rapidly between consecutive LAI points. This reflects the nonlinear relationship between canopy development and light interception described by the Beer–Lambert formulation used in the crop model, see Eqs. (2) and (3).

At early growth stages, when LAI is low, small variations in leaf area strongly affect the fraction of light intercepted by the crop. As a result, both fresh biomass and LAI increments show a stronger dependence on the growth-state variable in this region. Conversely, as the canopy develops and LAI increases, the intercepted light fraction progressively approaches saturation. For LAI values above approximately 5, the marginal effect of additional leaf area becomes limited, and both crop-related increments tend to vary more slowly with further increases in LAI. This explains why fewer breakpoints are required at higher LAI values to represent the crop response with sufficient accuracy.

Fresh biomass increment increased with PPFD over the entire LAI range, as expected from the larger amount of light available for photosynthesis. However, the response was clearly sublinear. The distance between consecutive PPFD curves decreased at higher light intensities, indicating that the marginal biomass gain associated with an additional increase in PPFD becomes progressively smaller. This behavior is consistent with the reduction in light use efficiency under high light intensity conditions. At maximum LAI, increasing PPFD from 100 to 500 μmol $m^{-2}$ $s^{-1}$ increased the fresh biomass increment by a factor of 3.6, whereas the lighting input increased by a factor of 5. Therefore, higher PPFD values still increase biomass production, but with decreasing marginal efficiency.

The LAI increment map showed a different response compared with the fresh biomass increment map. Although LAI increment also increased with PPFD, the differences among high PPFD levels were much less pronounced, especially above 300 μmol $m^{-2}$ $s^{-1}$. This stronger saturation is related to the dependence of leaf area expansion on specific leaf area, see Eq. (4). As reported in Appendix 6.2, SLA decreases as PPFD increases, meaning that plants grown under higher light intensity produce less leaf area per unit of accumulated dry matter. Consequently, the higher biomass accumulation obtained at high PPFD does not translate into a proportional increase in LAI.

These trends highlight the importance of including both fresh biomass and LAI increments in the surrogate formulation. Fresh biomass represents the productive output of the crop, while LAI acts as the growth-state variable governing canopy light interception and, indirectly, transpiration and energy demand. The different responses of these two variables to PPFD confirm that crop growth cannot be represented only through final biomass accumulation, especially when the surrogate model is used for dynamic lighting optimization over the crop cycle.

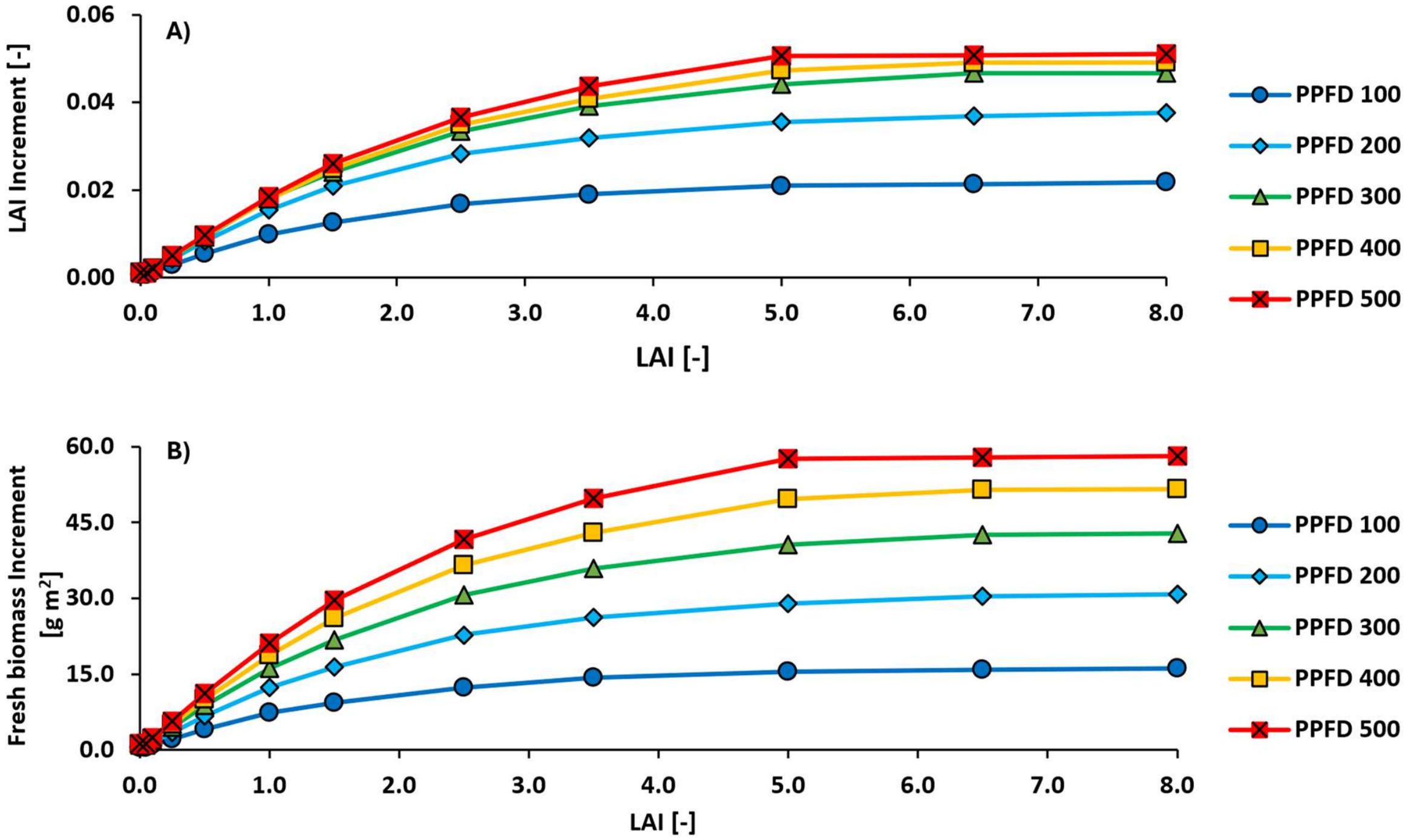

Figure 9: Piecewise crop growth relationships used in the MILP framework: hourly LAI increment (a) and hourly fresh biomass increment (b) as functions of crop growth state, represented by LAI, and selected PPFD level.

For the energy demand, this section reports only the electrical demand associated with heating and cooling, since these are the maps directly used in the MILP framework. The lighting electricity demand is not discussed in detail because it is mainly determined by the selected PPFD level and LED efficacy, with only limited nonlinearities in its formulation. The thermal demand maps before conversion into electricity are reported in Appendix 6.6. Heating and cooling electricity demands are expressed in Wh $m^{-2}$ of cultivated area to make the results more comparable and less dependent on the specific size of the analyzed vertical farm. Under the investigated winter conditions, outdoor temperature had a limited influence on the optimized schedules, since auxiliary electricity demand was dominated by cooling associated with LED heat release and crop latent loads. Variations in heat-pump COP were also partly moderated by part-load effects, making electricity price and crop developmental state the main optimization drivers. Therefore, outdoor temperature is not explicitly shown in the following energy-demand maps.

The heating demand includes both space heating and post-heating after dehumidification. As shown in Figure *10*, when the LEDs are off and LAI is low, the heating demand is only associated with space heating. This contribution remains limited because of the high insulation level of the vertical farm. As LAI increases, crop evapotranspiration also increases during the dark period. This increases the dehumidification requirement and, consequently, the post-heating demand. Once LAI reaches approximately 3.5, evapotranspiration approaches saturation and the heating demand tends to stabilize.

During the light period, the heat released by the LEDs reduces or eliminates the need for space heating. At low LAI, post-heating is also limited because the dehumidification requirement is low and can be covered within the cooling process. As the crop develops, evapotranspiration increases and the latent load becomes more relevant. As a result, heating demand appears mainly because post-heating is required after dehumidification. The LAI value at which this heating demand starts depends on PPFD. At lower PPFD levels, the heat released by the LEDs is smaller, so heating is required at earlier crop stages. At higher PPFD levels, the lighting system provides more heat to the growing chamber, delaying or reducing the heating requirement. For this reason, during the light period, heating demand decreases as PPFD increases.

The cooling demand includes both space cooling and dehumidification. When the LEDs are off and LAI is low, cooling demand is almost zero because there is no lighting heat to remove and crop evapotranspiration is still limited. As LAI increases, evapotranspiration increases and the cooling demand associated with dehumidification becomes relevant also during the dark period.

When the LEDs are on, cooling demand is already significant even at low LAI. In this case, the main contribution is the sensible heat released by the lighting system, which must be removed from the growing chamber. Therefore, cooling demand increases with PPFD. As LAI increases, the dehumidification load also increases because of higher crop evapotranspiration. However, the increase in total cooling demand with LAI is smoother than the increase observed for heating demand. This occurs because evapotranspiration increases the latent load, but it also produces a cooling effect that partially reduces the sensible cooling requirement.

At LAI values above approximately 3.5, the cooling demand tends to stabilize. In this region, evapotranspiration has reached saturation and the remaining cooling load is mainly associated with moisture removal and LED heat rejection.

Overall, the maps show that cooling is the dominant thermal service in the analyzed vertical farm. This is due to the combined need to remove the heat generated by the lighting system and to manage the latent load associated with crop evapotranspiration, especially at advanced crop development stages.

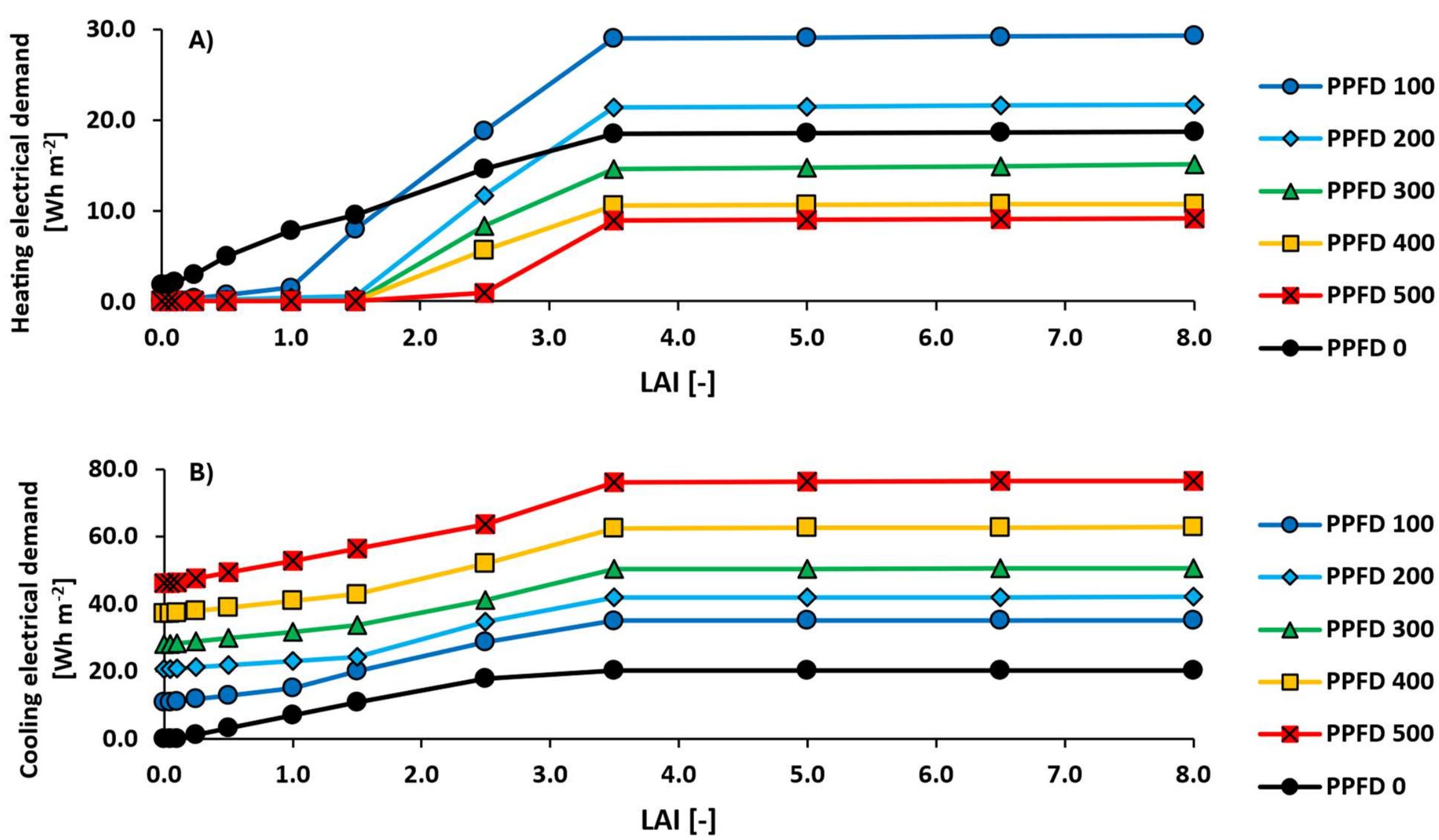

Figure 10: Piecewise heating and cooling demand relationships used in the MILP framework: hourly heating electrical demand (a) and hourly cooling electrical demand (b) as functions of crop growth state, represented by LAI, and selected PPFD level.

### 3.4. Optimization of Dynamic PPFD Schedules

The surrogate model described above was then used to analyze the selected case studies and identify cost-minimizing PPFD schedules as a function of crop growth stage and electricity price. The optimization aimed to reduce both the specific electric energy consumption, SEEC, defined as the electric energy required to produce 1 kg of fresh biomass, and the specific electricity cost, SC, defined as the electricity cost associated with the production of 1 kg of fresh biomass.

As described in the methodology, two fixed-lighting benchmark scenarios were considered. In both cases, the photoperiod was kept constant at 16 h $d^{-1}$, while PPFD was fixed at either 200 or 300 µmol $m^{-2}$ $s^{-1}$. Figure *11* compares the crop development and performance indicators of the two benchmark cases. The higher PPFD level allowed the crop to reach the target fresh weight of 250 g $plant^{-1}$ five days earlier than the 200 µmol $m^{-2}$ $s^{-1}$ scenario. However, this reduction in crop cycle duration was associated with a higher energy intensity. The 300 µmol $m^{-2}$ $s^{-1}$ benchmark achieved a SEEC of 9.25 kWh $kg^{-1}$, approximately 7% higher than the value obtained with the 200 µmol $m^{-2}$ $s^{-1}$ benchmark.

The impact on electricity cost was less pronounced than the increase in SEEC. Because of the temporal variability of electricity prices over the analyzed period, the higher energy consumption of the 300 µmol $m^{-2}$ $s^{-1}$ case resulted in an increase in SC of only 3.4% compared with the 200 µmol $m^{-2}$ $s^{-1}$ benchmark.

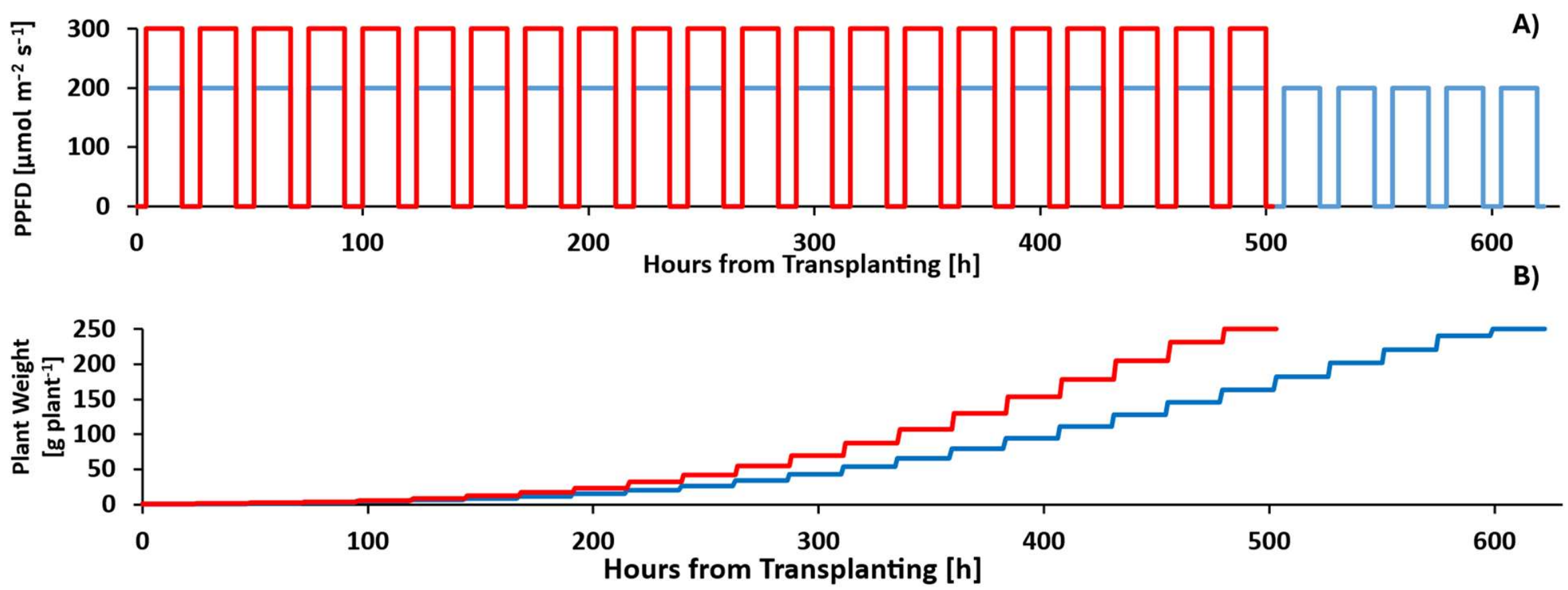

Figure 11: Comparison of the fixed-PPFD benchmark scenarios: imposed PPFD schedules (a) and corresponding crop growth profiles (b). Blue and red lines represent the 200 and 300 µmol m$^{-2}$ s$^{-1}$ scenarios, respectively.

### 3.4.1. Optimized PPFD Schedule with 24 h time aggregation

The first optimization was performed using the 24 h temporal aggregation. In this formulation, the optimizer could not modify the photoperiod adopted in the benchmark scenarios, which was fixed at 16 h d$^{-1}$, but it could adjust the PPFD applied during the daily light period according to electricity price and crop growth stage. The MILP problem was solved using Gurobi Optimizer. The number of variables and constraints depended on the crop cycle duration imposed by the corresponding benchmark scenario, namely 26 days for the 200 µmol m$^{-2}$ s$^{-1}$ case and 21 days for the 300 µmol m$^{-2}$ s$^{-1}$ case.

Although the 24 h formulation involved a relatively compact number of variables, the final biomass constraint represented a relevant limitation for the optimizer, since the selected PPFD schedule had to ensure that the target fresh weight was reached exactly within the prescribed crop cycle. Figure *12* compares the fixed benchmark PPFD profiles with the optimized 24 h schedules for the two case studies.

For the shorter 21-day cycle associated with the 300 µmol m$^{-2}$ s$^{-1}$ benchmark, the limited optimization horizon reduced the number of feasible scheduling alternatives. As shown in Figure *12*, the optimizer mainly reduced PPFD during the early growth stage. This choice exploited the higher light use efficiency observed at lower PPFD values without compromising the achievement of the target biomass at the end of the crop cycle. As a result, the optimized schedule reduced the SEEC by 4.2% compared with the corresponding fixed-PPFD benchmark.

A different behavior was observed for the 26-day cycle associated with the 200 µmol m$^{-2}$ s$^{-1}$ benchmark. The longer crop duration provided additional flexibility to redistribute light over the growth period. In this case, the optimizer selected relatively high PPFD values during the early stage, then reduced PPFD during the intermediate phase, and finally increased it again as LAI became larger. This strategy reflects the dependence of crop growth on intercepted light, which is strongly affected by canopy development. Providing sufficient light at the beginning of the cycle promoted the initial increase in LAI, while reducing PPFD during the intermediate stage limited energy consumption when the marginal benefit of additional light was lower. At later stages, when LAI was higher and the canopy intercepted a larger fraction of the incident light, the optimizer selected PPFD values above the benchmark level to compensate for the lower light input supplied during the intermediate phase.
Overall, the 24 h optimization reduced the SEEC by 9.9% compared with the 200 µmol $m^{-2}$ $s^{-1}$ benchmark. This result indicates that longer crop cycles provide greater scheduling flexibility and allow the optimizer to exploit the interaction between crop developmental stage, light interception, and energy demand more effectively.

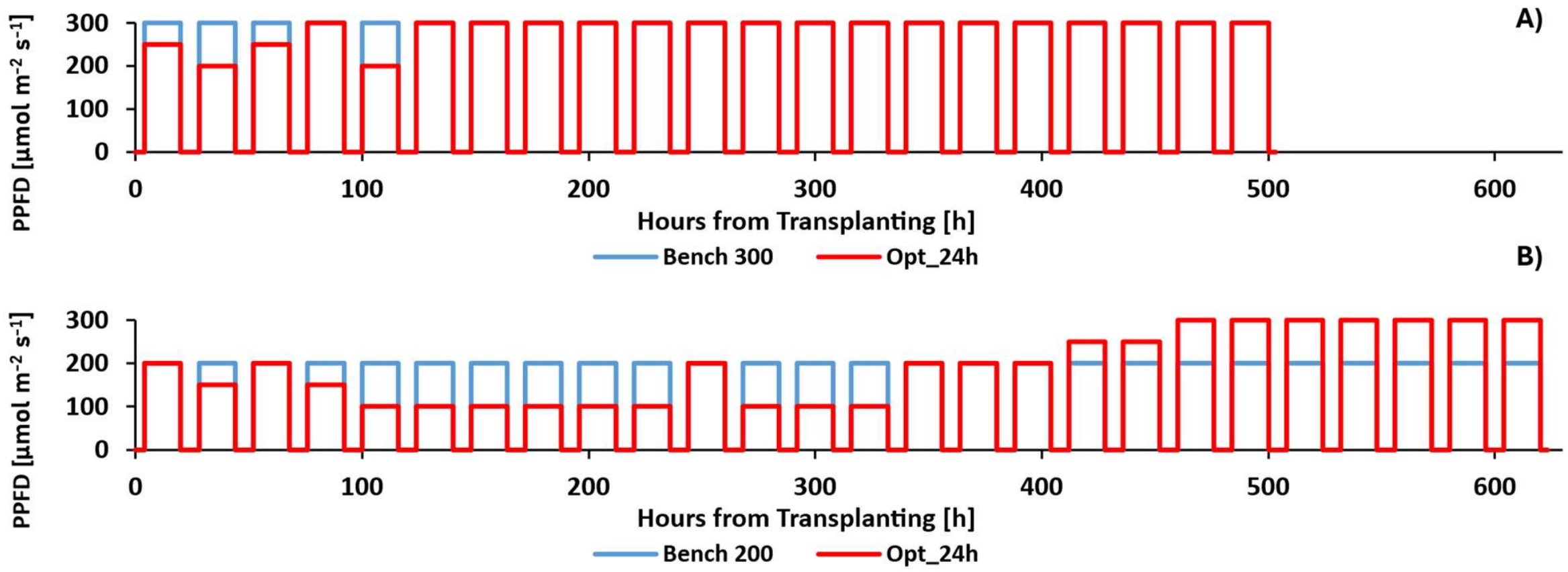

Figure 12: Comparison of PPFD schedules for the fixed-lighting benchmarks and the corresponding optimized solutions obtained with the 24 h temporal aggregation. Panels (a) and (b) refer to the 300 and 200 µmol $m^{-2}$ $s^{-1}$ benchmark scenarios, respectively.

To assess whether the approximation errors identified during the surrogate consistency analysis could accumulate over the crop cycle and affect the reliability of the optimized solution, the lighting schedule obtained from the MILP was applied as an external input to the nonlinear physical model. The resulting fresh matter production and cumulative energy demand were then compared with the corresponding predictions of the surrogate model, as shown in Figure *13*.
Overall, the surrogate model reproduced the trajectories generated by the physical model with satisfactory agreement over the entire cultivation cycle. At the end of the cultivation cycle, the relative deviations were approximately 4% for fresh matter and 6% for cumulative energy demand. The comparison was performed for the solution obtained with the 24 h temporal aggregation, since this formulation showed the largest approximation errors during the surrogate consistency assessment and therefore represents the most conservative validation case.
These results indicate that the local approximation errors introduced by the surrogate formulation do not propagate to an extent that materially alters the predicted crop development or energy consumption of the optimized schedule. The ex-post validation also supports the adopted 12-point LAI discretization as an appropriate compromise between approximation accuracy and computational tractability. A coarser LAI grid could increase cumulative state-transition errors and, consequently, reduce the reliability of the optimization results.
.

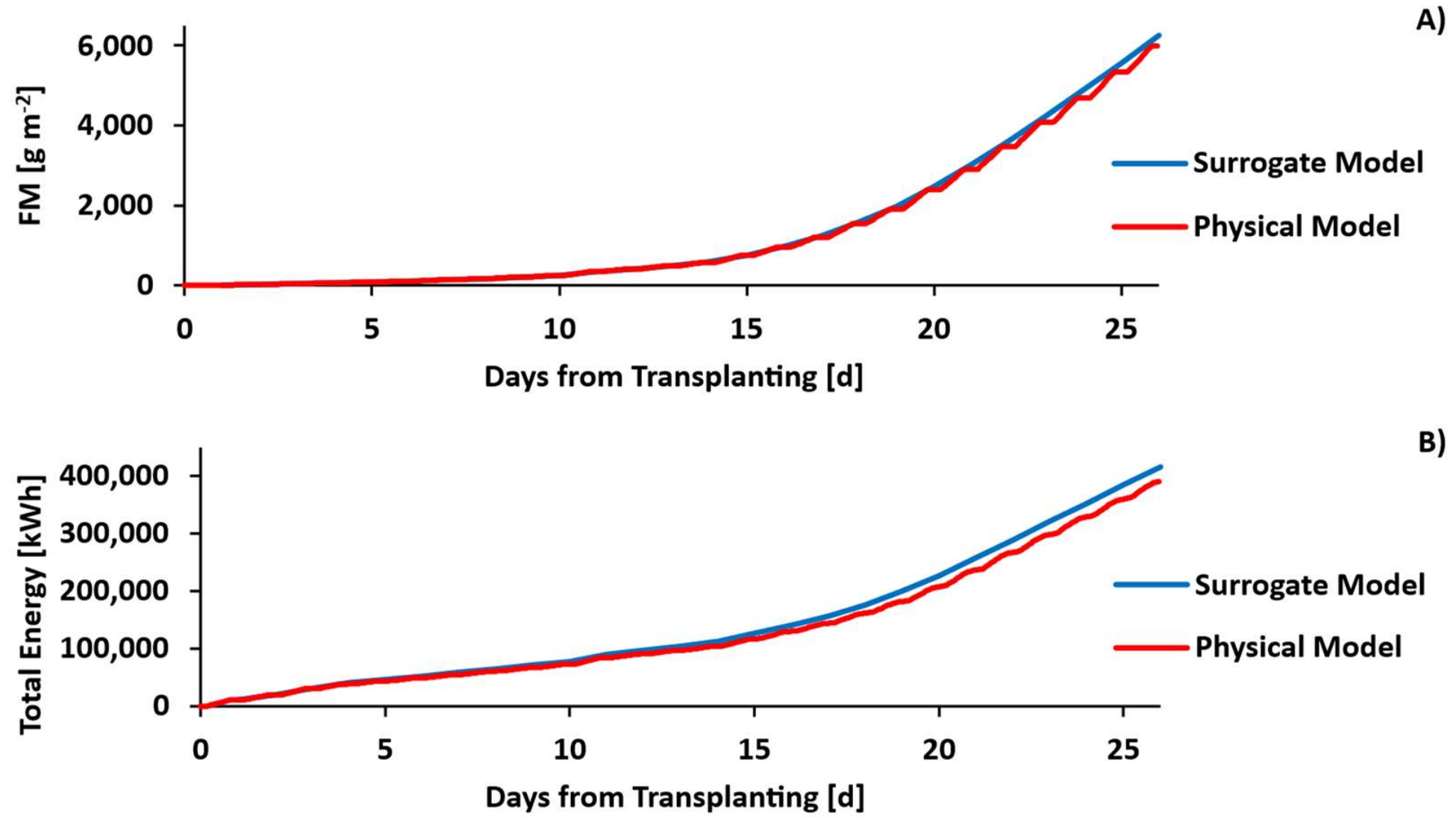

Figure 13: Ex-post validation of the optimized lighting schedule obtained with the 24 h temporal aggregation. Fresh matter production (a) and cumulative total energy demand (b) predicted by the surrogate model are compared with the corresponding outputs of the nonlinear physical model.

### 3.4.2. Optimized PPFD Schedule with 4 h and 1 h time aggregations

A more flexible optimization was then performed by allowing the MILP to adjust both PPFD and photoperiod. Compared with the 24 h formulation, this increased the scheduling flexibility available to the optimizer, since the daily light integral could be distributed over different photoperiods and PPFD levels. For brevity, only the results obtained with the 4 h temporal aggregation are reported in this section, as the optimized PPFD schedules obtained with the 1 h formulation showed only limited differences.

Figure *14* shows the optimized PPFD schedules for the two analyzed crop cycles and compares them with the corresponding fixed-lighting benchmarks. The general strategy identified with the 24 h aggregation was confirmed. In both case studies, the optimizer selected intermediate PPFD values during the early growth stages, lower PPFD values during the intermediate stages, and higher PPFD values toward the end of the crop cycle. This behavior reflects the interaction between light interception, crop developmental stage, and light use efficiency.

The possibility of adjusting the photoperiod further highlighted the advantage of distributing the daily light input over longer lighting periods at lower PPFD levels. Within the imposed DLI constraint, the optimizer generally preferred extended photoperiods combined with moderate PPFD values rather than shorter periods with high PPFD. This strategy was particularly relevant for the shorter 21-day crop cycle, where the optimizer reduced PPFD and increased the photoperiod in order to meet the target crop growth within the prescribed duration.

It is also noteworthy that, even with the finer temporal aggregation and time-varying electricity prices, PPFD values above 300 µmol $m^{-2}$ $s^{-1}$ were not selected as cost-effective solutions. This result is consistent with the reduction in light use efficiency observed at high PPFD levels, which limits the economic benefit of increasing light intensity beyond this range. Therefore, the optimizer tended to exploit longer lighting periods and moderate PPFD values rather than high-intensity lighting.

The 4 h formulation was also able to exploit electricity price fluctuations within the day. This is reflected by the PPFD variations observed within the same photoperiod in both scenarios. By shifting part of the light supply toward more favorable price intervals while respecting the agronomic constraints, the optimized schedules reduced the SEEC by 14.5% and 15.9% compared with the fixed-lighting benchmarks for the 26-day and 21-day crop cycles, respectively.

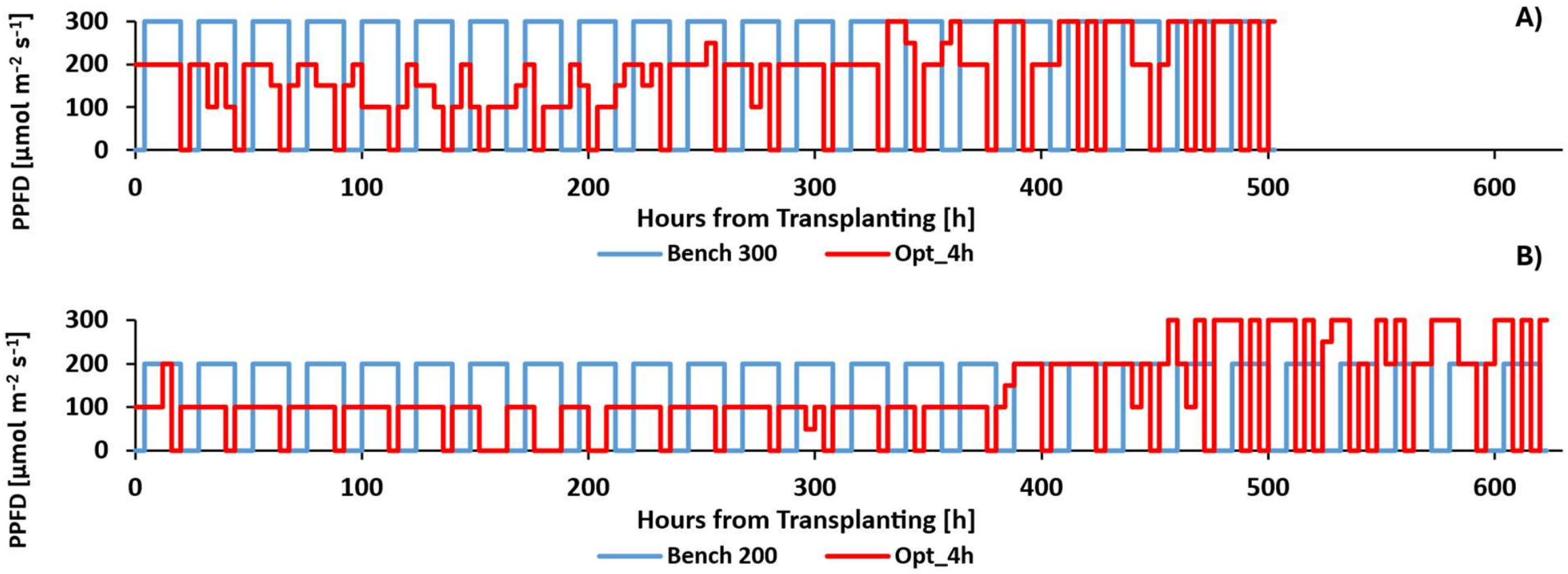

Figure 14: Comparison of PPFD schedules for the fixed-lighting benchmarks and the corresponding optimized solutions obtained with the 4 h temporal aggregation. Panels (a) and (b) refer to the 300 and 200 µmol $m^{-2}$ $s^{-1}$ benchmark scenarios, respectively.

### 3.4.3. Optimization overview

Using the piecewise linear surrogate model, the optimization framework identified several PPFD scheduling strategies that account for crop developmental stage and electricity price while ensuring that the target harvest weight is reached within the crop cycle duration prescribed by the user. The optimized schedules reduced both the specific electric energy consumption, SEEC, and the specific electricity cost, SC, compared with the fixed-lighting benchmark scenarios. The magnitude of these savings depended on both the imposed crop cycle duration and the temporal aggregation adopted in the MILP formulation, as summarized in Table *6*.

Finer temporal aggregations led to larger energy and cost reductions, because they provided the optimizer with greater flexibility in redistributing light over the crop cycle. The 24 h formulation, in which only the daily PPFD could be adjusted under a fixed photoperiod, achieved moderate reductions in SEEC and SC. Conversely, the 4 h and 1 h formulations allowed the optimizer to adjust both PPFD and photoperiod, thereby exploiting electricity price fluctuations more effectively. This is reflected by the fact that the cost savings obtained with these formulations were slightly higher than the corresponding energy savings. In other words, the optimizer did not only reduce the total energy required for production, but also shifted part of the lighting operation toward lower-price periods. The results also show that increasing the temporal resolution from 4 h to 1 h provided only marginal additional benefits. For both crop-cycle durations, the differences in SEEC and specific cost between the 4 h and 1 h formulations were negligible. This indicates that the 4 h aggregation already captured most of the economically relevant flexibility associated with electricity-price variability, photoperiod adjustment, and crop developmental stage. Therefore, for the investigated case study, a 4 h scheduling resolution appears sufficient to represent the main benefits of dynamic lighting management.
Overall, the best optimized schedules reduced SEEC from 8.63 to 7.37 kWh $kg^{-1}$ for the 26-day crop cycle and from 9.25 to 7.78 kWh $kg^{-1}$ for the 21-day crop cycle. These values correspond to energy savings of 14.6% and 15.9%, respectively. The reduction in SC was even higher, decreasing from 2.07 to 1.74 € $kg^{-1}$ for the 26-day cycle and from 2.14 to 1.76 € $kg^{-1}$ for the 21-day cycle, corresponding to cost savings of 16.0% and 17.8%. These results confirm that the proposed surrogate-based MILP framework can reduce both energy use and operating cost while satisfying the required crop production target. A detailed decomposition of the savings into energy-efficiency, HVAC, and price-shifting contributions could be considered in future applications, but it is outside the methodological scope of the present study.

Table 6: Comparison of benchmark and optimized lighting strategies in terms of specific electric energy consumption, specific electricity cost, and cost savings for the 26-day and 21-day crop cycles.

| **Crop Cycle Duration** | **26 days** | | | **21 days** | | |
|---|---|---|---|---|---|---|
| | **SEEC** | **SC** | **SC Savings** | **SEEC** | **SC** | **SC Savings** |
| Bench | 8.63 | 2.07 | [-] | 9.25 | 2.14 | [-] |
| Opt_24h | 7.78 | 1.90 | -8.1% | 8.86 | 2.06 | -3.7% |
| Opt_4h | 7.38 | 1.75 | -15.5% | 7.79 | 1.76 | -17.8% |
| Opt_1h | 7.37 | 1.74 | -16.0% | 7.78 | 1.76 | -17.8% |

## 4. Conclusions

This study presented a surrogate-based optimization framework to optimize lighting schedules in hydroponic vertical farming. One of the contributions is the definition of a general procedure to derive crop–energy relationships from response data and embed them into a tractable optimization framework. Although in this work the surrogate model was generated from synthetic data produced by a validated dynamic AMESim model, the same procedure can be applied to experimental or operational datasets from monitored vertical farms, provided that the relevant ranges of control variables, crop developmental stages, and energy outputs are adequately covered.
The framework was applied to lettuce cultivation in an industrial-scale vertical farm under dynamic electricity prices. The surrogate model described the dependence of crop growth and energy demand on PPFD, LAI, and outdoor temperature, allowing lighting schedules to be optimized while preserving the main biological and energy-related dependencies of the reference system. The results show that the proposed surrogate formulation reproduces the main trends of the physical model with sufficient accuracy and reduces both energy consumption and electricity cost compared with the selected fixed-lighting benchmarks, while meeting the modeled fresh-weight target and the imposed DLI and photoperiod constraints.
The main findings can be summarized as follows:

- The proposed procedure provides a general workflow to transform crop–energy response data into simplified relationships suitable for mathematical programming optimization, making the approach applicable not only to simulation-generated datasets but also to experimental datasets from real vertical farms.

- The surrogate model based on piecewise linearized relationships reproduced biomass and cumulative energy demand with good accuracy, especially for the 1 h formulation, with final-cycle CV(RMSE) values of 2.3% for biomass and 1.8% for energy, providing a tractable model for a MILP optimization framework.
- The optimized schedules reduced SEEC from 8.63 to 7.37 kWh $kg^{-1}$ for the 26-day cycle and from 9.25 to 7.78 kWh $kg^{-1}$ for the 21-day cycle, corresponding to energy savings of 14.6% and 15.9%.
- The specific electricity cost decreased from 2.07 to 1.74 € $kg^{-1}$ and from 2.14 to 1.76 € $kg^{-1}$ for the 26-day and 21-day cycles, respectively, corresponding to cost savings of 16.0% and 17.8%.
- PPFD values above 300 μmol $m^{-2}$ $s^{-1}$ were generally not selected, indicating that high-intensity lighting was not cost-effective under the investigated conditions because of the decreasing marginal light use efficiency.

Overall, the proposed workflow provides an accurate, computationally tractable, and broadly applicable framework for crop-energy optimization, while suggesting that optimized lighting schedules are governed primarily by crop developmental stage and should allocate more light to later growth phases, when canopy expansion enhances light interception and the productive use of supplied photons.

## 5. Appendix and supplementary data

### 5.1 Vertical Farm Design Input

Table A1: Main design and operating parameters adopted for the reference vertical farm configuration.

| Parameter | Value | Unit | Reference |
|---|---|---|---|
| VF Volume | 6,720 | $m^3$ | |
| VF Cultivated Area | 5,040 | $m^2$ | |
| Selected Crop | Lettuce | [-] | [20] |
| Photoperiod | 16 | h | [23] |
| $CO_2$ Concentration | 1,200 | ppm | [23] |
| Crop Density | 25 | Plants $m^{-2}$ | [23] |
| RH Setpoint | 75%/85% (light/dark) | [-] | [20] |
| Temperature | 24 | °C | [12] |

### 5.2 SLA-PPFD Relationship

The effect of incident PPFD on leaf area expansion per unit increase in dry biomass was derived from the experimental dataset reported by Carotti et al. [23] and is shown in Fig. A1. Higher PPFD values were associated with a lower increase in leaf area for the same dry matter accumulation, indicating a reduction in specific leaf area under high light conditions. In this study, only the relationship obtained at 24 °C was adopted, since this temperature was identified as the most favorable condition for lettuce growth in the vertical farm configuration considered.

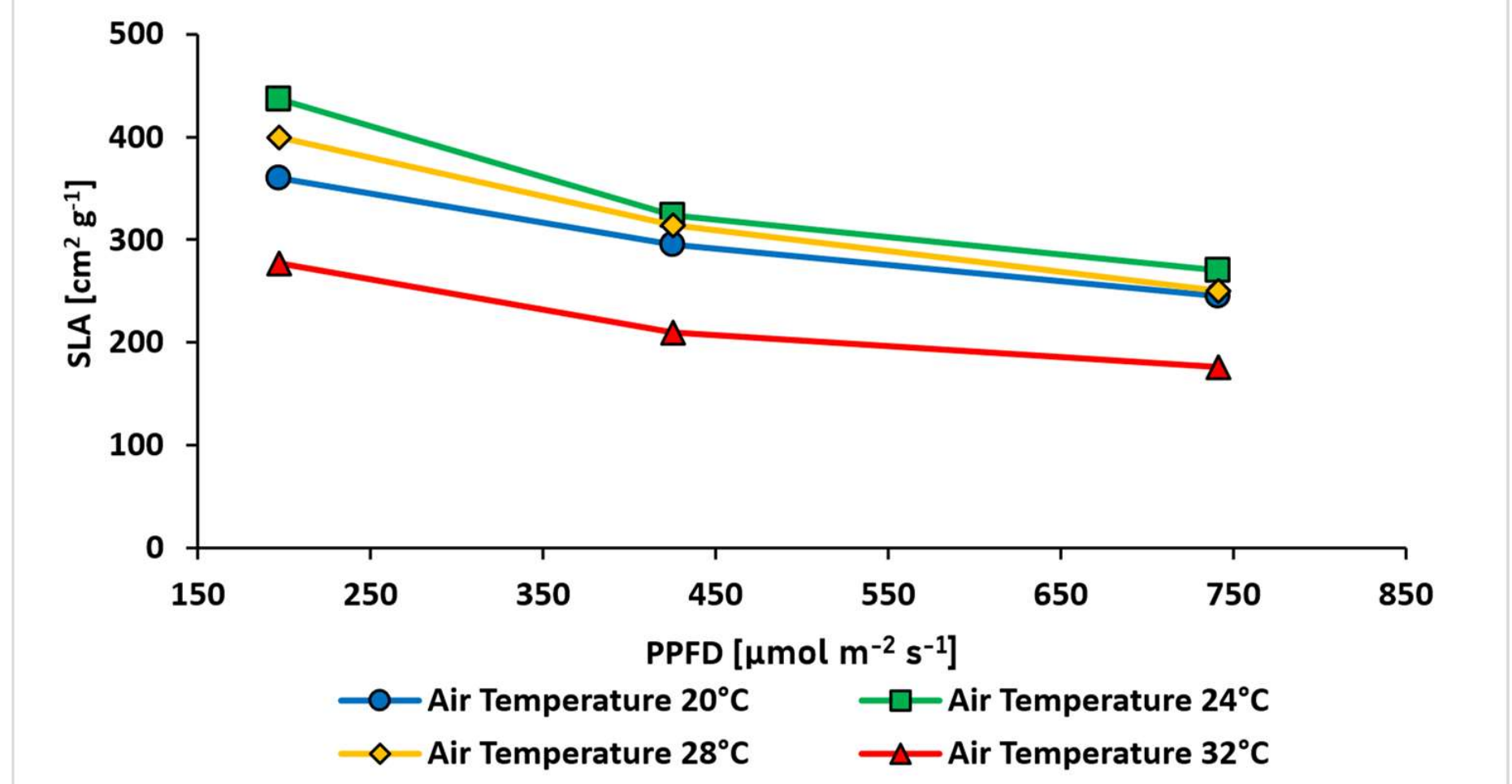

Figure A1: Specific leaf area as a function of PPFD and air temperature, derived from the experimental data used to parameterize the crop growth model.

## 5.3 Crop Transpiration Model

Crop transpiration was estimated at each timestep using the model proposed by Graamans et al. [26] The formulation is based on the canopy energy balance reported in Eq. (A1), where $R_{net}$ is the net radiation absorbed by the crop, H is the sensible heat flux exchanged between the leaf canopy and the surrounding air, and $\lambda E$ is the latent heat flux associated with water vapor transfer from the crop to the indoor environment. The sensible and latent heat fluxes are calculated according to Eqs. (A2) and (A3). In these equations, $T_s$ and $T_a$ denote the leaf surface temperature and indoor air temperature, respectively, while $r_s$ and $r_a$ represent the stomatal and aerodynamic resistances. Similarly, $X_s$ and $X_a$ are the vapor concentrations at the transpiring leaf surface and in the surrounding air. Net radiation was estimated by accounting for crop reflectance, assuming a constant reflection coefficient of 0.05 for photosynthetically active radiation (PAR).

$$R_{net} - H - \lambda E = 0 \quad \text{(A1)}$$

$$H = LAI \cdot \rho_a \cdot c_{p,a} \cdot \frac{T_s - T_a}{r_a} \quad \text{(A2)}$$

$$\lambda ET = \lambda \cdot LAI \cdot \frac{X_s - X_a}{r_s + r_a} \quad \text{(A3)}$$

The vapor concentration at the transpiring surface was estimated by linearizing the saturation condition around the indoor air temperature, following the first order Taylor expansion proposed by Monteith. In this formulation, $X_a^*$ represents the saturated vapor concentration at indoor air temperature, $\rho_a$ is the air density, $c_{p,a}$ is the specific heat capacity of air, $\lambda$ is the latent heat of vaporization, $\varepsilon$ is the slope of the saturation vapor pressure curve expressed in psychrometric form, and $T_s$ - $T_a$ is the temperature difference between the leaf surface and the surrounding air.

$$X_s \cong X^*{}_a + \frac{\rho_a \cdot c_{p,a}}{\lambda} \cdot \varepsilon \cdot (T_s - T_a) \quad \text{(A4)}$$

To solve the transpiration model, the leaf surface temperature must be evaluated at each timestep from the canopy energy balance. By rearranging the equations introduced above, the leaf surface temperature can be expressed as reported in Eq. (A5), which was implemented to close the system and compute the sensible and latent heat exchanges between the crop and the indoor air. The absolute humidity values required by the formulation were calculated directly within AMESim from the air mass balance of the growing room. The psychrometric coefficient ε, representing the local slope of the saturation relationship, was evaluated from indoor air temperature and vapor pressure according to Eq. (A6). The stomatal resistance $r_s$ was calculated according to the formulation proposed by Graamans et al. [26], reported in Eq. (A7), while the aerodynamic resistance $r_a$ and the LAI were assumed constant and equal to 100 s $m^{-1}$ and 3, respectively, following the recommendations provided in the reference study [26].

$$T_s(t) = T_a(t) + \frac{R_{net}(t) - LAI \cdot \lambda \cdot \frac{X^*{}_a(t) - X_a(t)}{r_s + r_a}}{LAI \cdot \rho_a \cdot c_{p,a} \cdot \left(\frac{1}{r_a} + \frac{\varepsilon(t)}{r_a + r_s(t)}\right)} \quad \text{(A5)}$$

$$\varepsilon(t) = \frac{\frac{4{,}098 \cdot p_{sat}(t)}{(T_a(t) + 2\quad .15)^2}}{0.067} \quad \text{(A6)}$$

$$r_s(t) = 60 \cdot \frac{1{,}500 + PPFD(t)}{200 + PPFD(t)} \quad \text{(A7)}$$

The computed leaf surface temperature was then used to evaluate the vapour concentration at the transpiring surface and, consequently, the crop transpiration rate. To account for canopy development during the growth cycle, the effective leaf area index defined in Eq. (A8) was introduced into Eq. (A3). This allowed the transpiration mass flow to vary with crop growth stage, producing a dynamic evapotranspiration profile rather than a constant rate over the cultivation period.

$$LAI_{eff}(t) = 3 - 0.2 \cdot \ln(1 + 5 \cdot (3 - LAI(t))) \quad \text{(A8)}$$

### 5.4 COP Correlations

The nominal second law efficiency was calculated using Eq. (A9), based on the heat pump nominal COP and the corresponding design outdoor and supply water temperatures. This value was then used as the reference efficiency for correcting the heat pump performance under off-design operating conditions. To improve the seasonal performance of the heating system, the operational hot water supply temperature was reduced by 5 °C with respect to the nominal design value, thereby decreasing the temperature lift required from the heat pump.

$$\eta_{II,nom} = \frac{COP_{nom}}{\frac{T_{hw,nom} + 273.15}{T_{hw,nom} - T_{ext,nom}}} \quad \text{(A9)}$$

Part load effects were included for both heating and cooling operation. The heating part load factor and the cooling part load factor were derived from [28] and [29], respectively, and implemented as correction factors applied to the nominal performance of the corresponding unit. Their profiles as a function of load ratio are reported in Fig. A2. This allowed the model to account for the efficiency degradation associated with off design operation and partial load conditions.

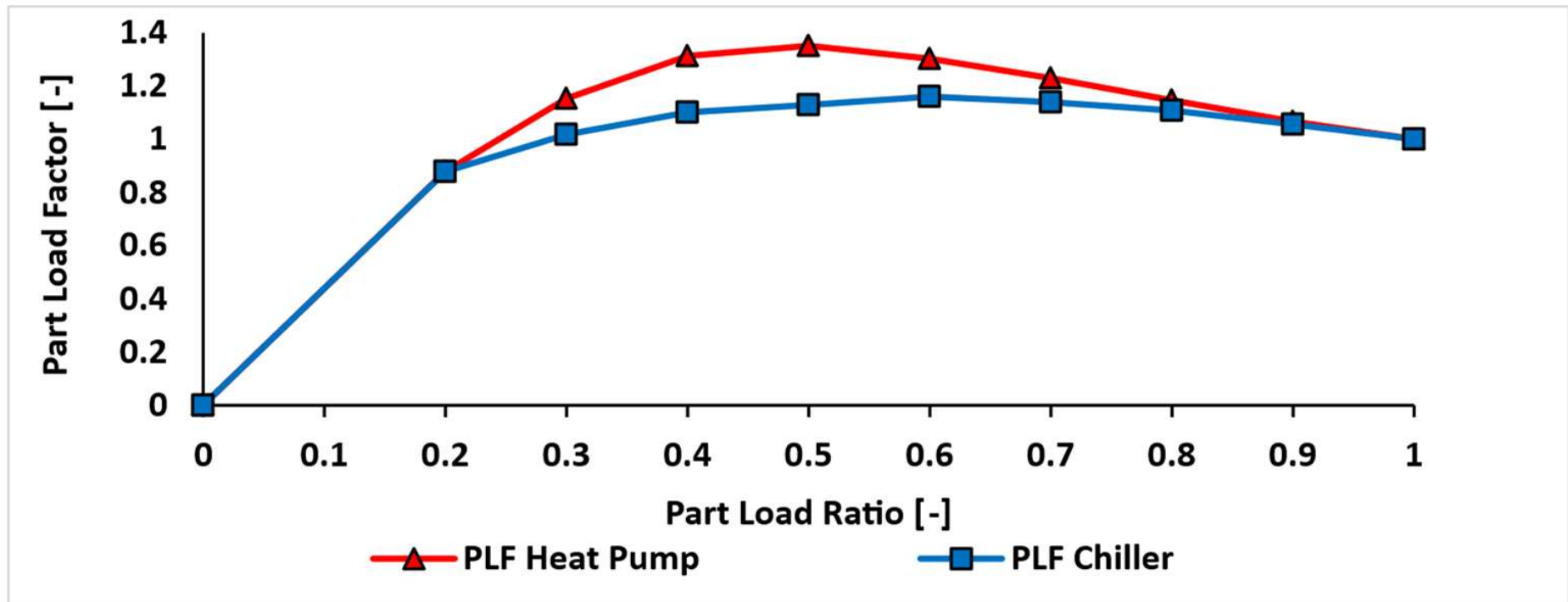

Figure A2: Part load factor curves adopted for the heat pump and chiller as a function of the part load ratio.

As noted above, outdoor conditions were not assumed to directly affect chiller performance in the present study, since the simulated period refers to winter operation and the outdoor air temperature remains below the lower bound of the compressor performance map. Under this assumption, the chiller EER varies only as a function of part load operation. Therefore, the resulting EER profile follows the cooling part load factor, scaled by the nominal EER.

Conversely, heat pump performance is strongly affected by outdoor air temperature. Variations in $T_{ext}$ influence both the second law efficiency and the maximum heating capacity available at each timestep, thereby affecting the resulting COP under off design conditions. For this reason, the COP was evaluated as a function of both outdoor air temperature and heating part load ratio. The resulting COP profiles are reported in Fig. A3, highlighting the combined effect of source temperature and partial load operation on heat pump performance.

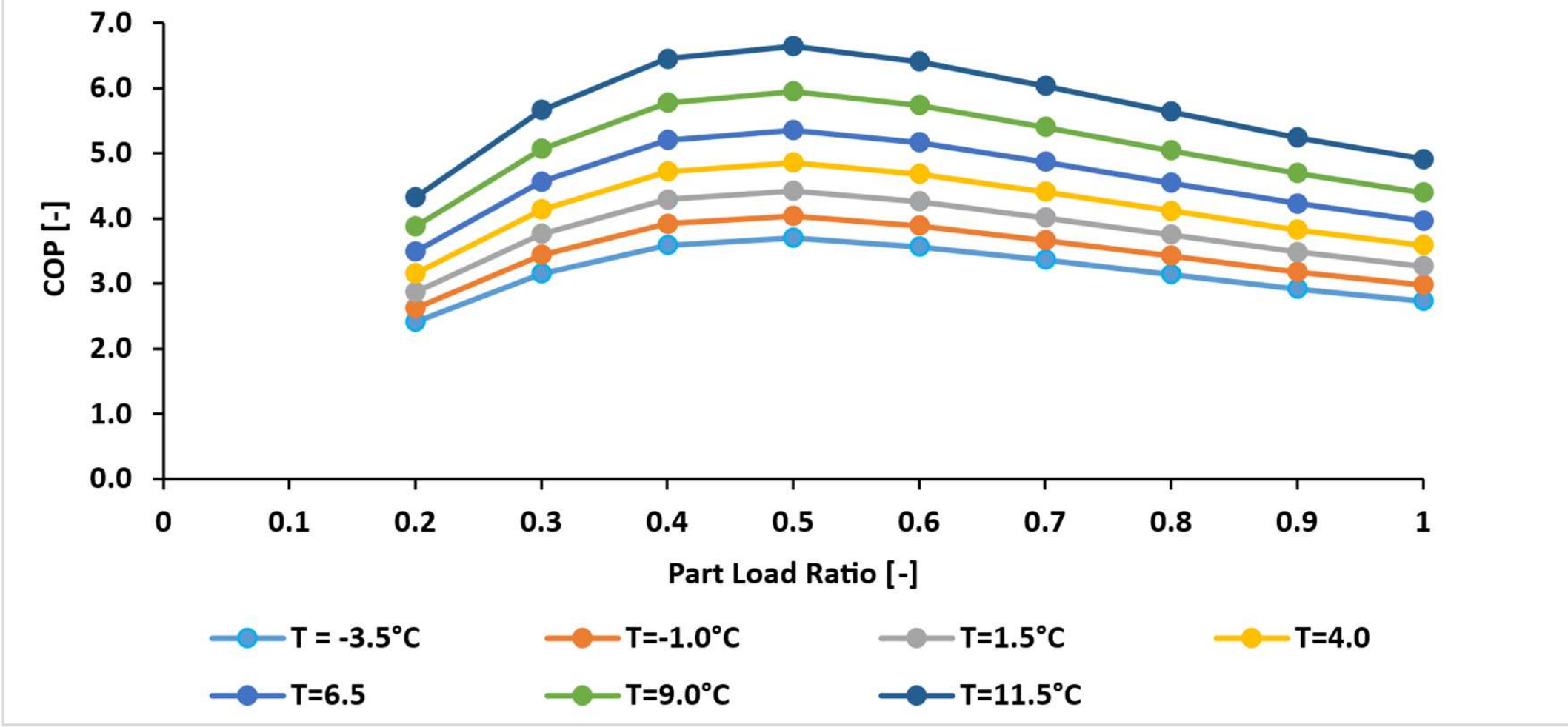

Figure A3: Heat pump COP as a function of part load ratio and outdoor air temperature, as obtained from the adopted off-design performance correlation.

## 5.5 Climatic Data

The external temperature profile, used by the model to assess energy consumption and heat pump performance, was derived from [31] for a representative winter period and is reported in Fig. A5.

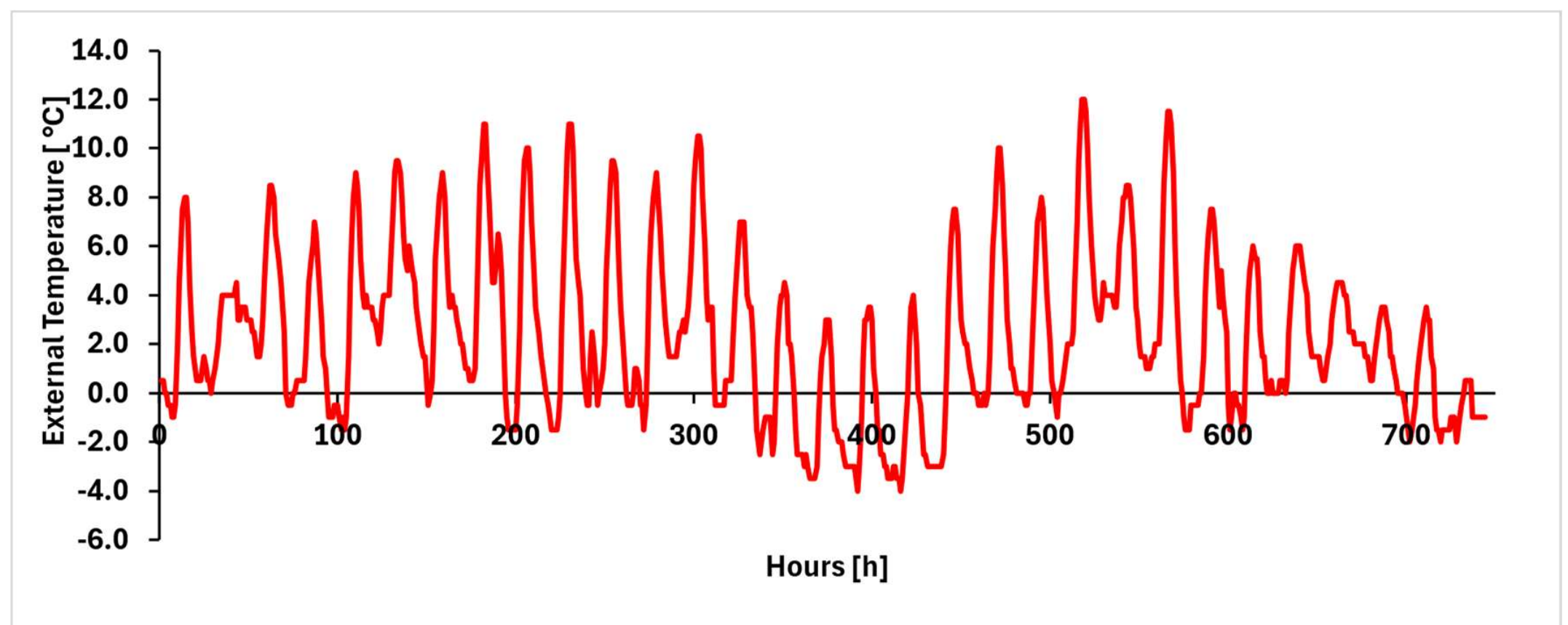

Figure A4: Outdoor air temperature profile adopted for the selected winter period in the case study.

### 5.6 Surrogate model: growth and energy demand piecewise maps

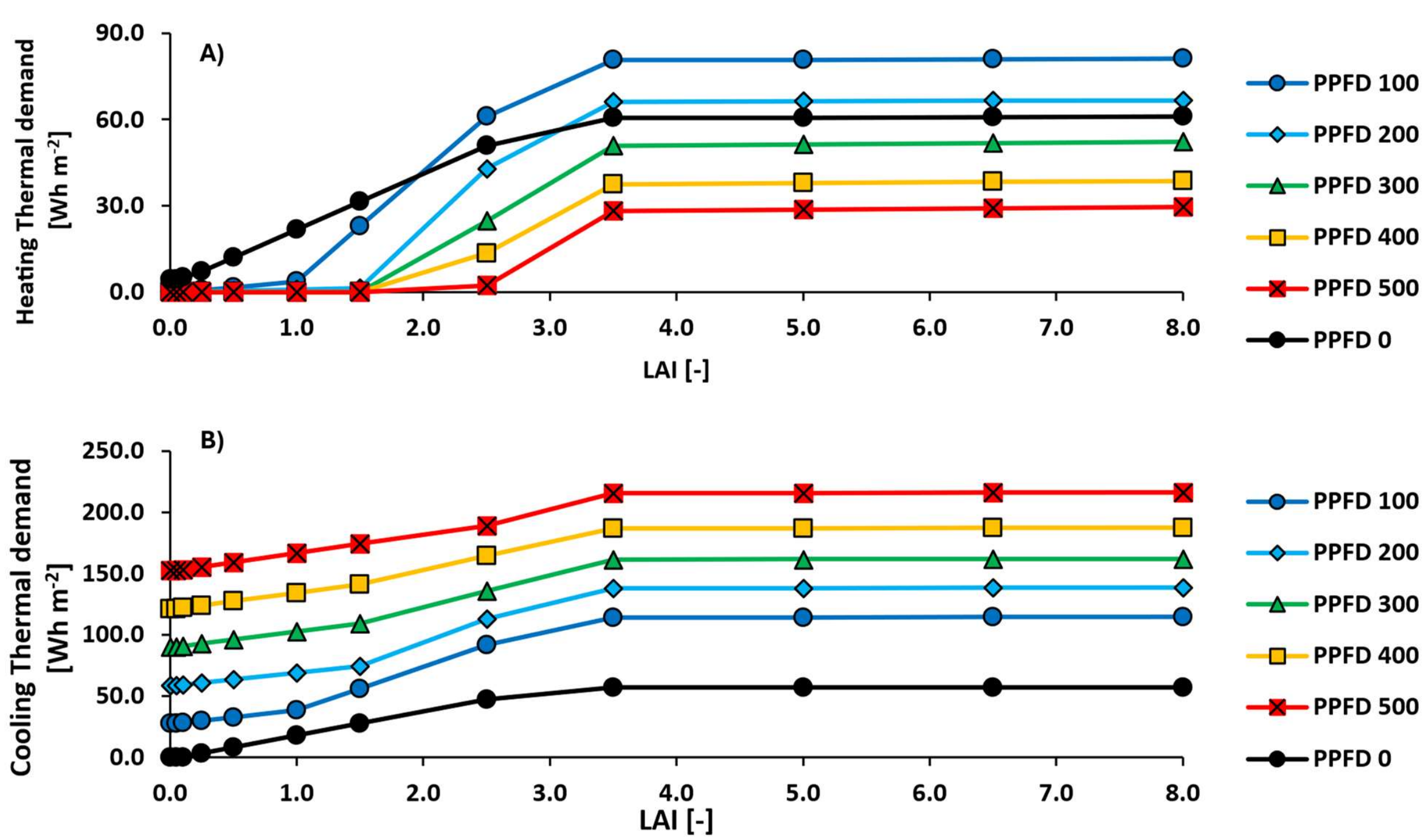

Figure A5: Heating and cooling thermal demand maps as functions of crop growth state, represented by LAI, and selected PPFD level. Panels (a) and (b) report the heating and cooling thermal demand, respectively.

### 5.7 MILP Characteristics

The MILP was implemented in Python and solved using Gurobi Optimizer 12.0.3 on a Windows Server 2022 machine equipped with an Intel Xeon Gold 5120 CPU at 2.20 GHz, using up to eight threads. The largest instance, corresponding to a 26-day cultivation cycle with 1 h temporal resolution, included 60,531 variables, of which 4,368 were binary and 52,416 were SOS2 interpolation weights, together with 9,415 linear constraints and 4,368 SOS2 sets. The model was solved with a 24 h time limit and a target MIP gap of 1%, using MIPFocus = 1, Heuristics = 0.5, Presolve = 2, Cuts = 2, and NumericFocus = 1. A feasible warm start derived from the corresponding 4 h solution was supplied to improve incumbent identification.

For the most computationally demanding instances, the solver reached the prescribed time limit before closing the optimality gap. During the final stage of the computation, no relevant improvement in the incumbent solution was observed, while the best bound remained nearly stationary. The best feasible schedule available at termination was therefore retained, although global optimality was not formally certified. Conversely, smaller instances showed substantially stronger convergence. In particular, the 21-day crop cycle with 24 h temporal aggregation was solved with a final MIP gap of 0.94%, providing a near-optimality certificate for that case.

**Declaration of generative AI and AI Usage**

During the preparation of this work, the authors used ChatGPT 5.5 in order to improve the readability of the manuscript. After using this tool/service, the authors reviewed and edited the content as needed and take full responsibility for the content of the published article.